\documentclass[twocolumn,showkeys,preprintnumbers,amsmath,amssymb]{revtex4}
\usepackage{epsfig,latexsym}
\usepackage{graphicx}
\usepackage{dcolumn}
\usepackage{bm}
\begin{document}

\title[Luminance Gradients \& Luminosity Perception]{Gradient Representations and the Perception of Luminosity}

\author{Matthias S. Keil}%
\email{matsATcvc.uab.es; threequarksATyahoo.com; AT=@}
\affiliation{%
Basic Psychology Department,
Faculty for Psychology,
University of Barcelona (UB),
Passeig de la Vall d'Hebron 171,
E-08035 Barcelona (Spain)}%
%
%
%
\date{\today}
\begin{abstract}
The neuronal mechanisms that serve to distinguish between light-emitting
and light reflecting objects are largely unknown.  It has been suggested that
luminosity perception implements a separate pathway in the visual system,
such that luminosity constitutes an independent perceptual feature.
Recently, a psychophysical study was conducted to address the question
whether luminosity has a feature status or not.  However, the results
of this study lend support
to the hypothesis that luminance gradients are instead a perceptual
feature.  Here, I show how the perception of luminosity can emerge
from a previously proposed neuronal architecture for generating
representations of luminance gradients.
\end{abstract}

%
\maketitle

\def\Cite#1{\cite{#1}}	

\def\pics{./}

\def\eq[#1]{equation~\ref{#1}}
\def\Eq[#1]{Equation~\ref{#1}}
\def\eqs[#1]{equations~\ref{#1}}
\def\Eqs[#1]{Equations~\ref{#1}}

\def\Caption[#1][#2][#3]{\caption{\label{#1}\small{{\bf #2.} #3}}}

\def\fig[#1]{figure~\ref{#1}}
\def\Fig[#1]{Figure~\ref{#1}}
\def\sec[#1]{section~\ref{#1}}

\def\chessrampx{chess ramp}
\def\chessramp{{\chessrampx }}

\def\glowx{\emph{glow}}
\def\scramx{\emph{scrambled}}
\def\halox{\emph{halo}}
\def\controlx{\emph{control}}
\def\glow{{\glowx} }
\def\scram{{\scramx} }
\def\halo{{\halox} }
\def\control{{\controlx} }
\def\s{{ }}

\def\para{\\}			
\section{Introduction}
\label{sect:intro}

%
Under daylight illumination conditions, looking at a television or computer screen
rarely produces the sensation that displayed items are light-emitting, although each
pixel of the screen emits light (\cite{ZavagnoCaputo2001}, with references).\\
But to perceive objects as being luminous, it is not necessary to have a physically
source of light emission.  Halos were used by artists since a long time
as a means to create luminosity effects in their paintings
(\cite{ZavagnoCaputo2001}, with references). 
When a region is painted with a halo surrounding it, then one perceives this region
with enhanced brightness, or even as glowing, without physical light emission being
present.
Thus, the perception of glow \emph{can} be evoked on (light reflecting) paper
or canvas, and text or pictures being displayed on a (light emitting) computer
screen are \emph{not necessarily} being perceived as luminous.\\
In other situations perception and physics are not divergent.
For example, the sun is always perceived as light emitting,
and so are stars at night.  In such situations, the strong contrast between
light sources and background may provide the key factor to the perception
of luminosity \Cite{BonatoGilchrist1994,BonatoGilchrist1999}.\\
A recent fMRI study has identified a region in the brain which seems to be
associated with  the perception of luminosity \Cite{LeonardsTrosciankoLazeyrasIbanez05}.
In this study, different configurations of the glare effect display
(\cite{Kennedy76,BressanMingollaSpillmannWatanabe1997,Zavagno1999};
\fig[setups], top row) were presented to human observers.
The results of the study were indicative to that luminosity might constitute
a perceptual feature much like contrast, orientation, motion, or faces.
The question about whether luminosity is a perceptual feature or not
motivated a corresponding psychophysical study \Cite{CorreaniSamuelLeonards06}.
The study was based on the idea that perceptual features are distinguished from other object
properties by being processed in a more efficient way.  This means that
visual features consume less attentional resources than non-features
\Cite{JosephChunNakayama1997}, what is reflected in, for example,
``pop out'' effects.  A visual search paradigm such as the one used
in the study of \cite{CorreaniSamuelLeonards06}, therefore can serve
to distinguish features from non-features.\\
Unexpectedly, the results of Correani \emph{at al.} are compatible with that
\emph{luminance gradients} instead of luminosity are a visual feature.
Several authors have already formulated the hypothesis that luminance
gradients are involved in the perception of luminosity
\Cite{Kennedy76,Zavagno1999,ZavagnoCaputo2001,ZavagnoCaputo2005}, as there
is evidence that luminance gradients can influence lightness perception
under certain circumstances.\\
I therefore asked whether a recently proposed theory for the perception of
luminance gradients (``gradient system'') could account for the just-described
observations.  The gradient system has been successful in quantitatively predicting
available data on Mach bands \Cite{gradVisRes06}.  It furthermore provided an
account for Chevreul's illusion in terms of luminance gradients \Cite{MatsNC06},
and in addition is capable of real-world image processing.\\
In this work I will show how spatial configurations of luminance gradients
can interact to produce the perception of luminosity in the absence of
physical illuminants.  The results presented here also contribute
to the further understanding of how luminance gradients interact with
lightness computations and brightness perception, respectively.
Specifically, representations of luminance gradients provide a
straightforward explanation of ``self-luminous grays''
\Cite{ZavagnoCaputo2001,ZavagnoCaputo2005}, and why it is that
perception of luminosity is independent from lightness anchoring.
%
%
\begin{figure}[ht!]
	\begin{center}
		\scalebox{0.45}{\includegraphics{\pics/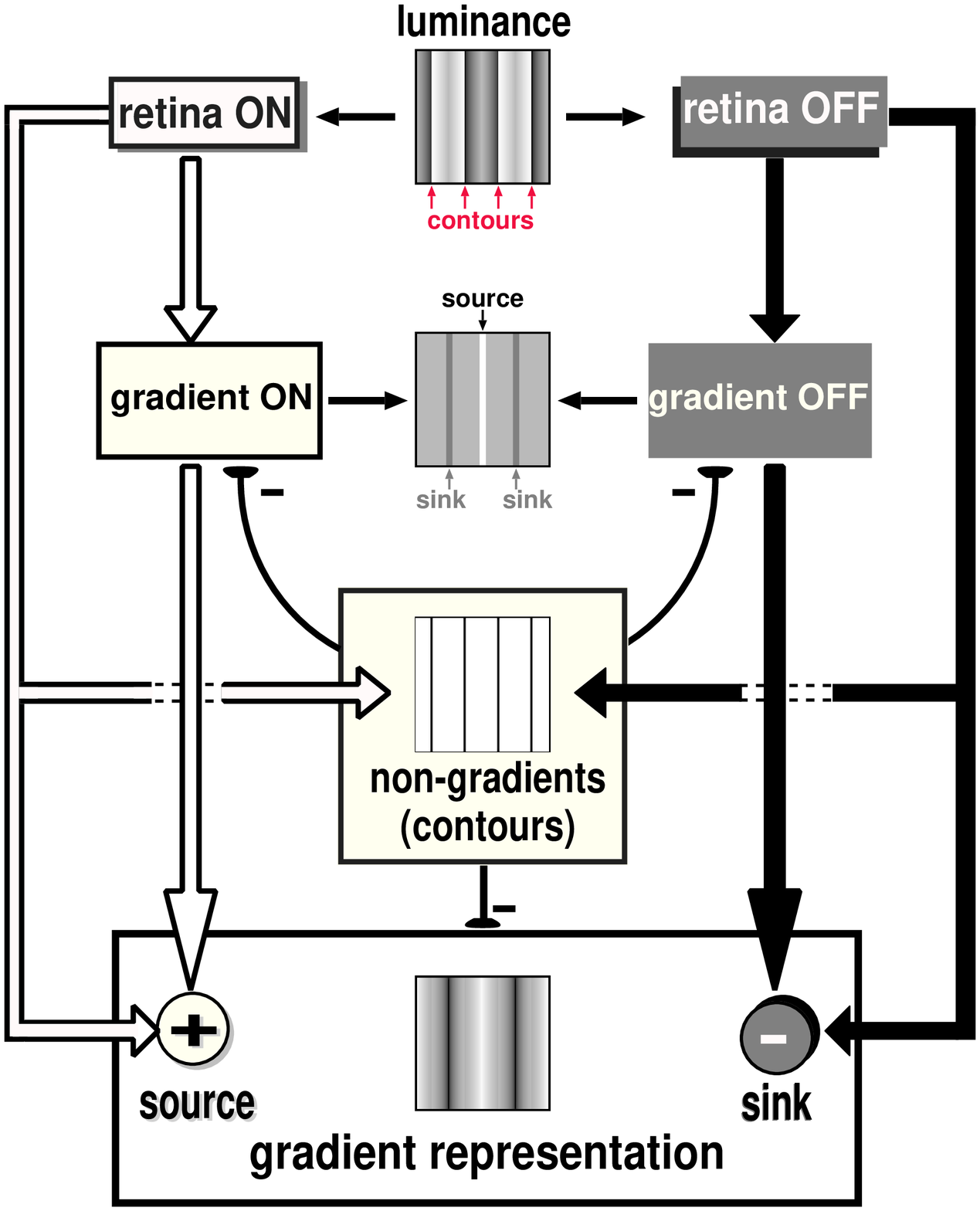}}
	\end{center}
	\Caption[howitworks][Functional overview over the gradient system][{%
	A \emph{notched square wave grating} (or briefly ``notch grating'') is
	used for illustration of the processing stages.  A notch grating is a square wave
	with notches being centered at each luminance step, and luminance decays
	(for the bright stairs) and increases linearly (for the dark stairs),
	respectively, to a common luminance level (the luminance profile is
	shown in \fig[sourcessinks]).  This means that the faint lines centered at
	each step have the same intensity value, yet they are perceived with different brightness.
	See \sec[subsect:howitworks] for a detailed explanation of the processing stages.}]
\end{figure}
\begin{figure}
	\begin{center}
		\scalebox{0.50}{\includegraphics{\pics/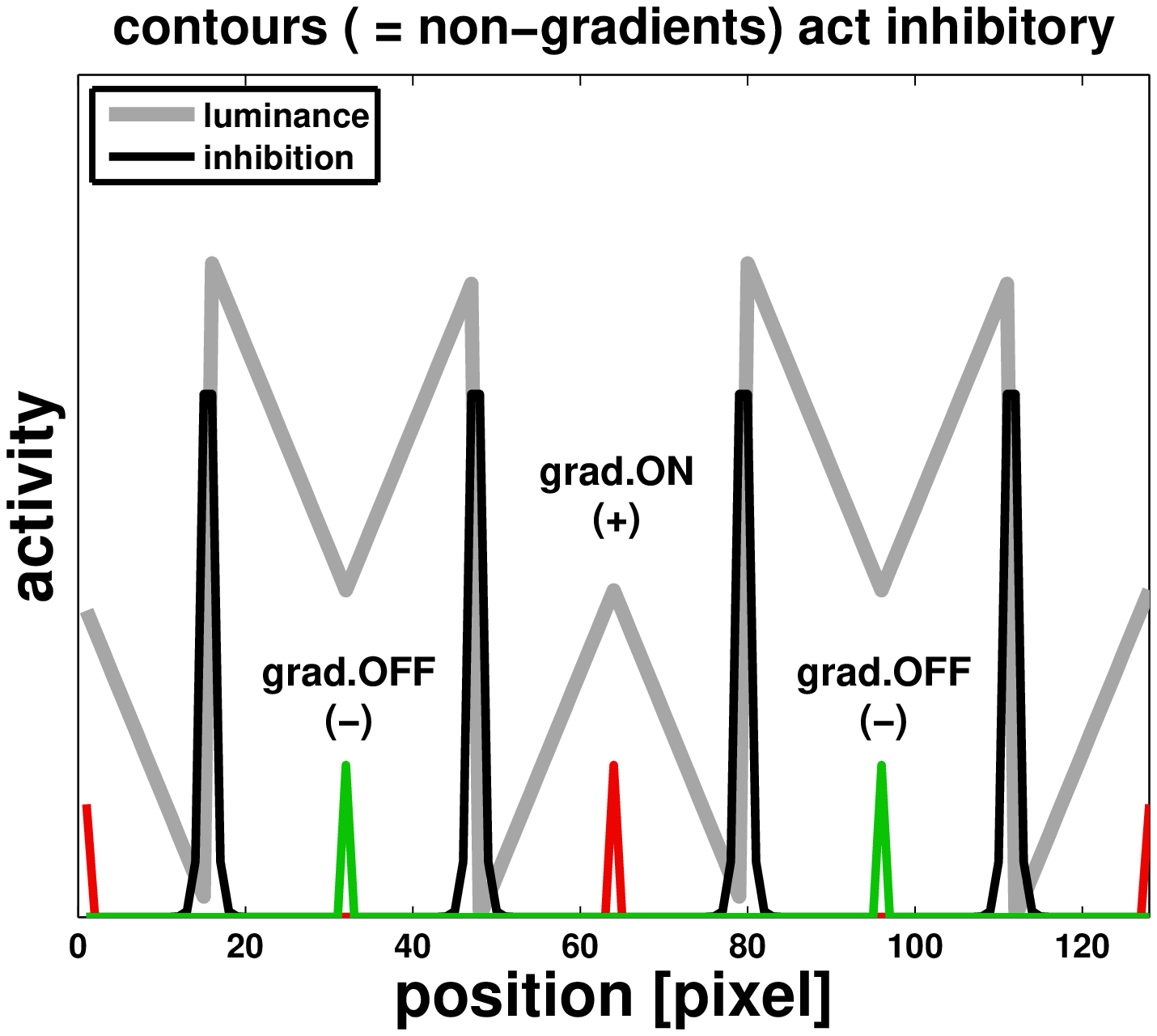}}
	\end{center}
	\Caption[sourcessinks][Gradient enhancement and contours ($=$ non-gradients)][{%
	The \emph{notched square wave grating} (or briefly ``notch grating'', legend label
	``luminance'') is a periodic spatial pattern composed of step-like changes
	and linear luminance gradients (the notches).  Contours are detected at
	the step-like changes in luminance.  Contours are related to surface
	processing, and thus should be suppressed in gradient representations.
	In the gradient system, the suppression is executed by contours acting
	inhibitory (see legend label).  This \emph{non-gradient} inhibition
	leaves just those activity patterns in retinal channels which correspond
	to smooth changes in luminance (\emph{gradient ON} activity ``$(+)$'',
	and \emph{gradient OFF}  activity ``$(-)$''; c.f. \fig[howitworks]).  During
	the creation of a gradient representation, gradient ON and OFF patterns
	eventually act as \emph{sources} and \emph{sinks}, respectively.  Depending
	on whether sources and sinks correspond to a linear luminance gradient
	(as shown here) or not, a gradient has to be explicitly be generated or not,
	respectively (see \fig[WaveProfiles]).}]
\end{figure}
\begin{figure}
	  {\bf \large (a)}\scalebox{0.45}{\includegraphics{\pics/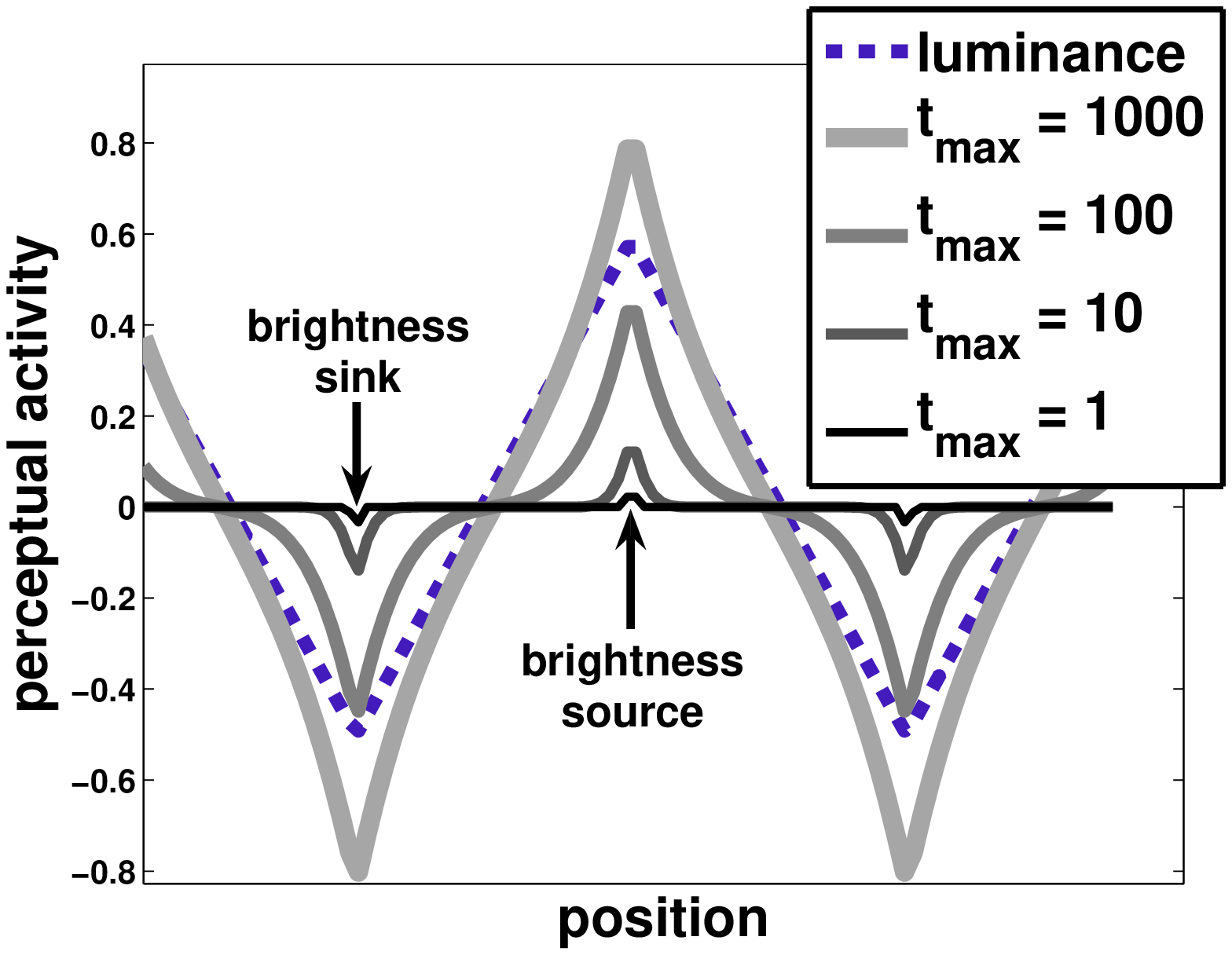}}\\
	  {\bf \large (b)}\scalebox{0.45}{\includegraphics{\pics/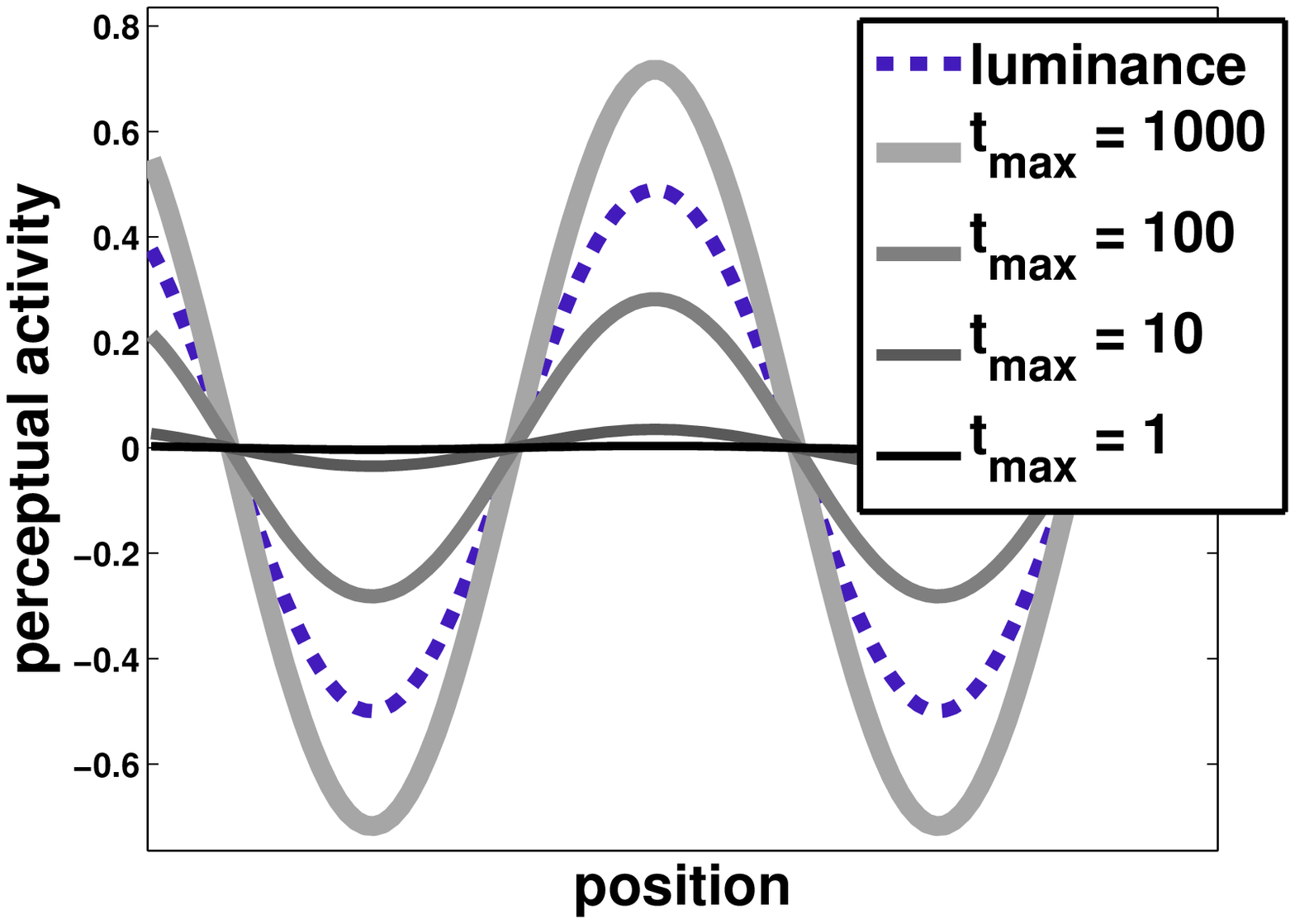}}
	\Caption[WaveProfiles][Linear and nonlinear luminance gradients][{%
  	(Both plots show activity profiles of two-dimensional representations at different
	times -- see legend).  	Linear luminance gradients \emph{(a)} are processed by the
	gradient system differently to nonlinear gradients \emph{(b)}.  In the former case,
	an activity gradient has to be explicitly generated by lateral spread of activity
	between a brightness \emph{source} and a brightness \emph{sink} (a a matter of fact,
	a brightness source is equivalent to a darkness sink, and a brightness sink is equivalent
	to a darkness source).  Sources and sinks may be localized activity patterns as in
	\emph{(a)} (where they are indicated by arrows), but be also spatially more extended
	as in \emph{(b)} (i.e., for nonlinear luminance gradients).  In the initial gradient
	representation, sources and sinks of nonlinear gradients are just a low-activity
	version of their final representation.  Thus, representations for nonlinear luminance
	gradients are produced by only amplifying the initial activity pattern,
	similar to a standing wave with increasing amplitude.  The corresponding luminance
	displays and their gradient representations in 2-D are shown in \fig[MoreExamples].}]
\end{figure}
%
%
%
\section{Introducing the gradient system}
\label{sect:gradsystem}
%
%
This section provides an overview over important characteristics of the gradient system.
A more detailed description of it, as well as its formal definition, can be found in
\cite{MatsNC06} and \cite{gradVisRes06}.

%
%
\subsection{Motivation}
\label{subsect:motivation}
%
%
The original motivation for proposing representations of luminance gradients
was that they are of different utility for object recognition.  It is known,
for example, that they may aid to \emph{(i)} recover three-dimensional information 
to compute surface shape (shape from shading, e.g. \cite{MingollaTodd86,Ramachandran88}),
\emph{(ii)} to resolve the three-dimensional layout of visual scenes
(e.g. \cite{KeKnMaBu96,BlojKerstenHulbert99}), \emph{(iii)} to identify material
properties of object surfaces (e.g., mat versus glossy), and are therefore
complementary to lightness computations (lightness is associated with surface
representations).\\
In situations, however, it may happen that luminance gradients rather would
interfere with the goal of generating invariant surface representations, and
thus disrupt lightness constancy.
(Invariant surface representations are mandatory for robust object recognition).  
In natural scenes, specular highlights, cast shadows, and slow illumination
gradients are often superimposed on object surfaces.  In such cases,
luminance gradients must be suppressed in surface representations for
establishing lightness constancy.  Nevertheless, it has been demonstrated
recently that humans use cues such as shadows, shading and highlights
for segregation of object surfaces \Cite{FowlkesMartinMalik07}.  Thus,
lightness constancy implies discounting ``gradient features'' on the
one hand, yet on the other hand they are used by humans to achieve a
more reliable segregation of figural regions from the background.\\ 
Taken together, luminance gradients contain different information,
which cannot be interpreted by bottom-up mechanisms.  Without
segregating them from surfaces, surface representations
would vary as a function of illumination conditions and scene layout.
Notice that such a merged representation would necessitate segregation
anyway, as  lightness constancy is not interrupted by specular highlights
\Cite{ToddFarleyMingolla04}, and human object recognition seems to work
reliably for most illumination conditions and scenes.\\
Having separate representations for surfaces and gradients, however,
has the advantage that gradient representations could be dynamically
linked to lightness computations \Cite{vdMalsburg95}.  In this way, mechanisms
for object recognition could utilize or suppress corresponding information,
what could contribute to increase robustness.\\
%
\subsection{How it works}
\label{subsect:howitworks}
%
%
The gradient systems is a hypothetical neuronal circuit, and its main processing
stages are shown in \fig[howitworks]
(see also figure $1$ in \cite{gradVisRes06}).  The retina constitutes two pathways,
which are related to brightness (``ON-channel''), and darkness (``OFF-channel''),
respectively.  A high-resolution boundary map is produced by processing information
from both channels\footnote{In \cite{MatsNC06}, and \cite{gradVisRes06}, a simplified
circuit is used to this end, which detects contours without using orientation-selective
operators.}.  ``High-resolution'' is to say that only the finest scale is considered.
At a cortical level, boundary maps are usually regarded as demarcating surface
representations thus defining surface shape.  Because contours define surfaces,
but not gradients, they are referred to as \emph{non-gradients} within the
gradient system.  Non-gradients act always inhibitory (\fig[sourcessinks]).\\
In the first step of gradient processing, gradients are enhanced 
by suppressing ON- and OFF-activity at non-gradient positions.
The result of this process can be conceived as ``retinal activity
maps with erased contours'' (``gradient ON'' \& ``gradient OFF''
in \fig[howitworks]).\\
In the second step, retinal ON-activity and gradient ON-activity
provide excitatory input to the site labeled by ``$+$'' in \fig[howitworks].
Analogously, OFF-activity from retina and gradients act inhibitory
on the site labeled by ``$-$''\footnote{For the sake of simplicity,
ON- and OFF-channels interact directly for generating the gradient representations.
The channels are distinguished by their respective sign, where information
from the ON-channel has a positive sign, and information from the
OFF-channel corresponds to negative values.  See \cite{MatsNC06},
p. 882 for more details.}.  Excitation and inhibition is
tonic or \emph{clamped}, what means that activity is actively
generated at ``$+$'' and ``$-$.  In addition, activity spreads laterally:
Activity values with positive sign from ``$+$'', and negative values
from ``$-$''.  Silent (or shunting) inhibition (reversal potential
equals resting potential that is zero) exerted by non-gradient
features during activity propagation
quickly suppresses boundaries, while at the same time
gradient activity is further enhanced.  As a consequence,
\emph{sources} and \emph{sinks} are dynamically created\footnote{%
In \cite{MatsNC06} and \cite{gradVisRes06}, sources and sinks were
simply defined as retinal ON plus gradient ON, and retinal OFF plus
gradient OFF, respectively.  These ``static'' sources and sinks
are identical to the dynamically created ones if an input image
contains only smooth changes in luminance, but no step-like changes.}.
Because of lateral propagation processes, activity gradients will
eventually form between sources and sinks (but see \fig[WaveProfiles]).
This latter process is referred to as
\emph{clamped diffusion}\Cite{MatsNC06,gradVisRes06}.\\
Silent non-gradient inhibition imposes a further important
constraint on the creation of gradient representations: Gradients cannot
spread beyond a surface over which they were originally superimposed.
This constraint also implies that activity gradients could form between
a source and a site of active non-gradient inhibition, but also between
a sink and a site of active non-gradient inhibition.  Such behavior
occurs, for example, with the \emph{notched square wave grating}
(``notch grating'', figures \ref{howitworks}, \ref{sourcessinks},
\ref{interactions} \& \ref{interactions_profiles}).\\
The gradient system generates representations of linear luminance gradients
by lateral propagation of activity between a brightness source and a brightness
sink.  At equilibrium, an activity gradient has formed between source and sink
(\fig[WaveProfiles]\emph{a}).\\
On the other hand, nonlinear luminance gradients, such as sine wave gratings,
need not to be explicitly created as it is the case with linear gradients.
Rather,  the initial activity pattern is only amplified (\fig[WaveProfiles]\emph{b}).
Notice that representations of linear and nonlinear gradients are generated by the
same mechanism, that is clamped diffusion.\\
Summarizing, there are three components which influence in the generation
of gradient representations.  \emph{(i)}  Brightness sources are created
from the retinal ON-channel, and their activity is related to ``brightness''.
Brightness sources constantly generate activity with positive sign.  This
activity propagates laterally. \emph{(ii)} Brightness sinks are the
counterpart of brightness
sources, and originate from the retinal OFF-channel.  Brightness sinks are
are identical with darkness sources, because they generate negative-valued
activity. By the same arguments are brightness sources identical with
darkness sinks.  If a stimulus only contains luminance gradients, then only
brightness sources and brightness sinks will influence in the formation
of gradient representations, where activity gradients will form between
sources and sinks (or between brightness sources and darkness sources).
\emph{(iii)} If the stimulus, however, contains surface structures,
silent non-gradient inhibition will be evoked, which strictly speaking
acts as an activity drain for both brightness and darkness activity. 
Non-gradient inhibition, however, does not actively generate activity.
To avoid name clashes, the terms ``sources'' and ``sinks'' are
exclusively reserved for brightness sources and brightness sinks,
respectively.  The term ``drain'' is used to refer to activity
dissipation because of non-gradient inhibition.
\emph{(iv)} Representations of linear and nonlinear luminance
gradients are generated by the same mechanism (clamped diffusion).
%
%
%
\section{Material and methods}
\label{sect:methods}
%
%
All results were generated with the implementation of the gradient system as described in \cite{MatsNC06},
and \cite{gradVisRes06}, respectively.  All parameter values and numerical methods were also the same
for the present study as before.  Simulations were carried out with a Matlab environment (R2006b)
on a Linux workstation.  If not otherwise stated, gradient representations were evaluated at
$t_{max}=1000$ iterations.
For the figures \ref{ChessK}, \ref{displays_ChessK} and \ref{CentralSquareFixed}, gradient activity
was averaged across the positions of the central square of the input (see \fig[GlareParade]).
Spatial averaging was carried out separately for brightness (i.e., positive values) and darkness (i.e.,
negative values), respectively.  In both of the last figures, the figure label ``perceptual activity''
means that the absolute value of average darkness was subtracted from average brightness at each data
point.  In \fig[CentralSquareVaries], only brightness activity is shown, as the first
data point of all curves (corresponding to luminance zero of the central square) gave $-0.0022$  for computing
\emph{average brightness} minus \emph{average darkness}, and the abscissa was scaled logarithmically.  Each
of the images in \fig[setups] and \ref{GradsGlowParade} showing gradient representations were normalized
individually in order to improve the visualization.   For the figures \ref{GlowSketch} and \ref{FluorentSketch},
the image size was $256 \times 256$ pixels.  For the rest of the simulations, luminance displays were of size
$128 \times 128$ pixel.  Luminance values were in the range from $0$ (black) to $1$ (white).
%
\begin{figure*}[ht!]
	\begin{center}
		\scalebox{0.5}{\includegraphics{\pics/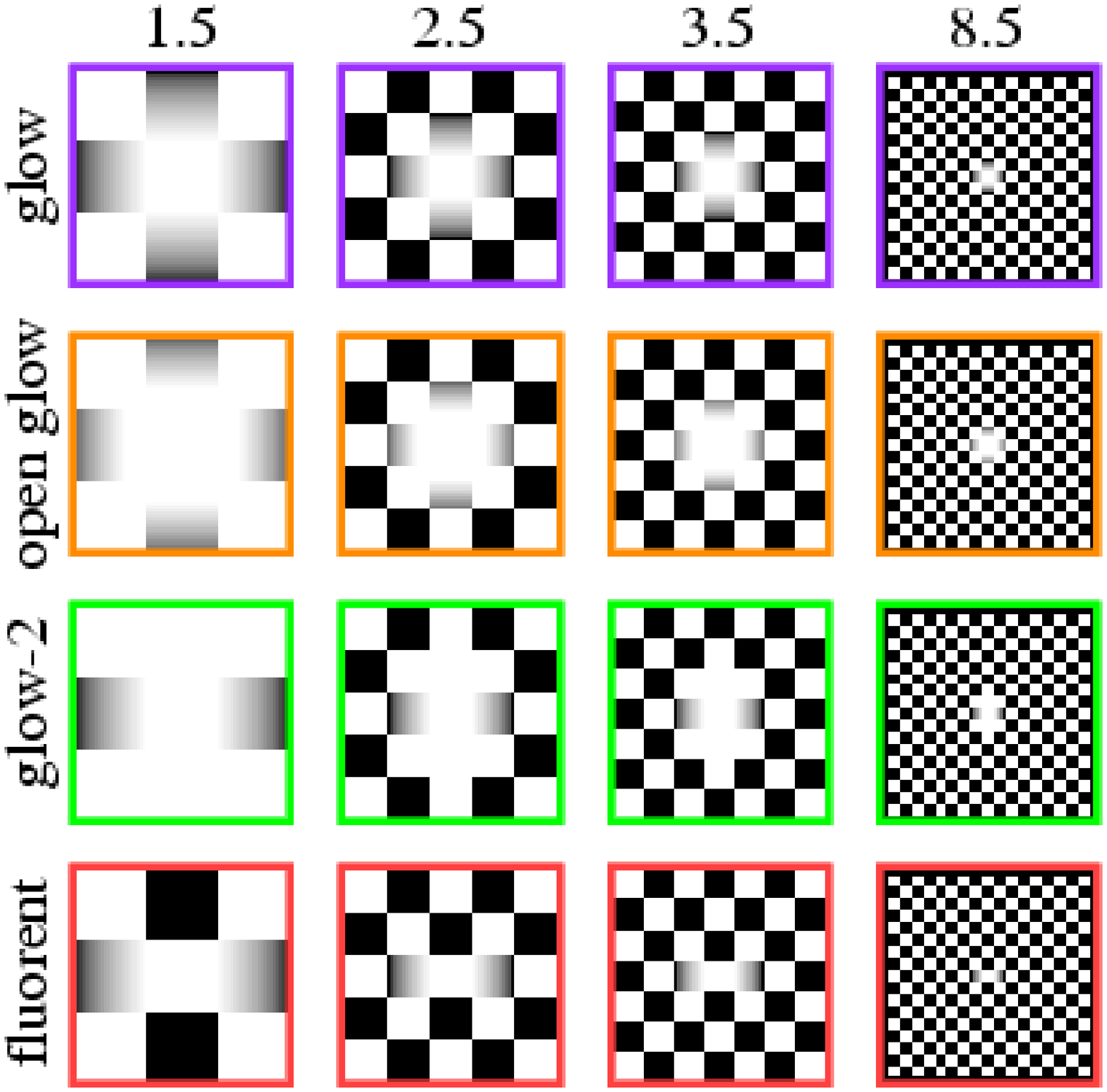}}
	\end{center}
	\Caption[GlareParade][A glow parade][{The figure shows three modifications
	(``open glow'', ``glow-2'', and ``fluorent'') of the original glare effect
	display shown in the first row (``glow'') at four spatial frequencies of
	the chessboard carrier (number denoting columns correspond to cycles per
	image).  Gradient representation of these images are shown in \fig[GradsGlowParade].
	The center square of each image appears as being light-emitting, albeit
	the strength of the effect seems to depend on display configuration and
	spatial frequency.}]
\end{figure*}
\begin{figure}[h!]
	\begin{center}
		\scalebox{0.65}{\includegraphics{\pics/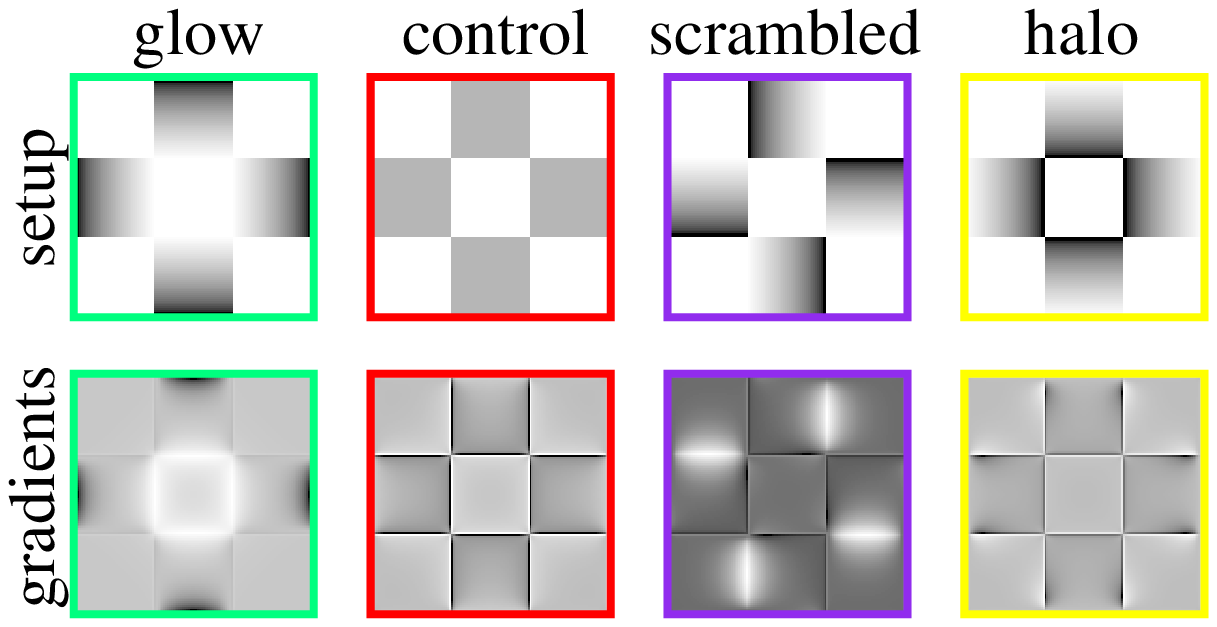}}
	\end{center}
	\Caption[setups][Setups][{The top row defines the four setups
	\glowx, ¸\controlx, \scramx, and \halo as employed in the present
	study (setups are distinguished by \emph{italic} letters).
	The definition follows the luminance displays as they were
	introduced in the study of \cite{CorreaniSamuelLeonards06}.
	Self-luminosity is only perceived for the \glow setup, but
	in none of the other cases.  Notice, however, the light Mach bands
	at the white end of the ramp for the \scram setup (the Mach
	bands are reproduced as white lines in the gradient
	representations).  The bottom row shows the corresponding
	gradient representations.}]
\end{figure}
%
\section{Results of simulations}
\label{sect:results}
%
%
\subsection{The glare effect}
In the present study, the glare effect display was systematically modified and corresponding responses
of the gradient system were studied.  The original glare effect (as introduced in \cite{Zavagno1999})
is shown in the first image of \fig[GlareParade].  It consists of a chessboard image (\emph{carrier}),
in which four black squares were substituted by luminance ramps (\emph{inducer squares}).  The white
field of the chessboard which is surrounded by the luminance ramps is the \emph{target square} or
\emph{central square}.  Notice that the ramps are linear gradients.
Depending on the spatial arrangement of the luminance ramps with
respect to the central square, it is perceived as being light-emitting in the \glow setup
(all images in \fig[GlareParade], first image in \fig[setups]).  If the luminance ramps
are arranged according to the \scram setup or the \halo setup (\fig[setups]), then one
cannot observe any brightness enhancement of the central square
\Cite{LeonardsTrosciankoLazeyrasIbanez05,CorreaniSamuelLeonards06}.  Similarly,
no brightness enhancement occurs in the \control configuration, where the four inducer
squares are set to a homogeneous luminance value - the mean value of a inducer square.
%
\begin{figure}
	  {\bf \large (a)}\scalebox{0.45}{\includegraphics{\pics/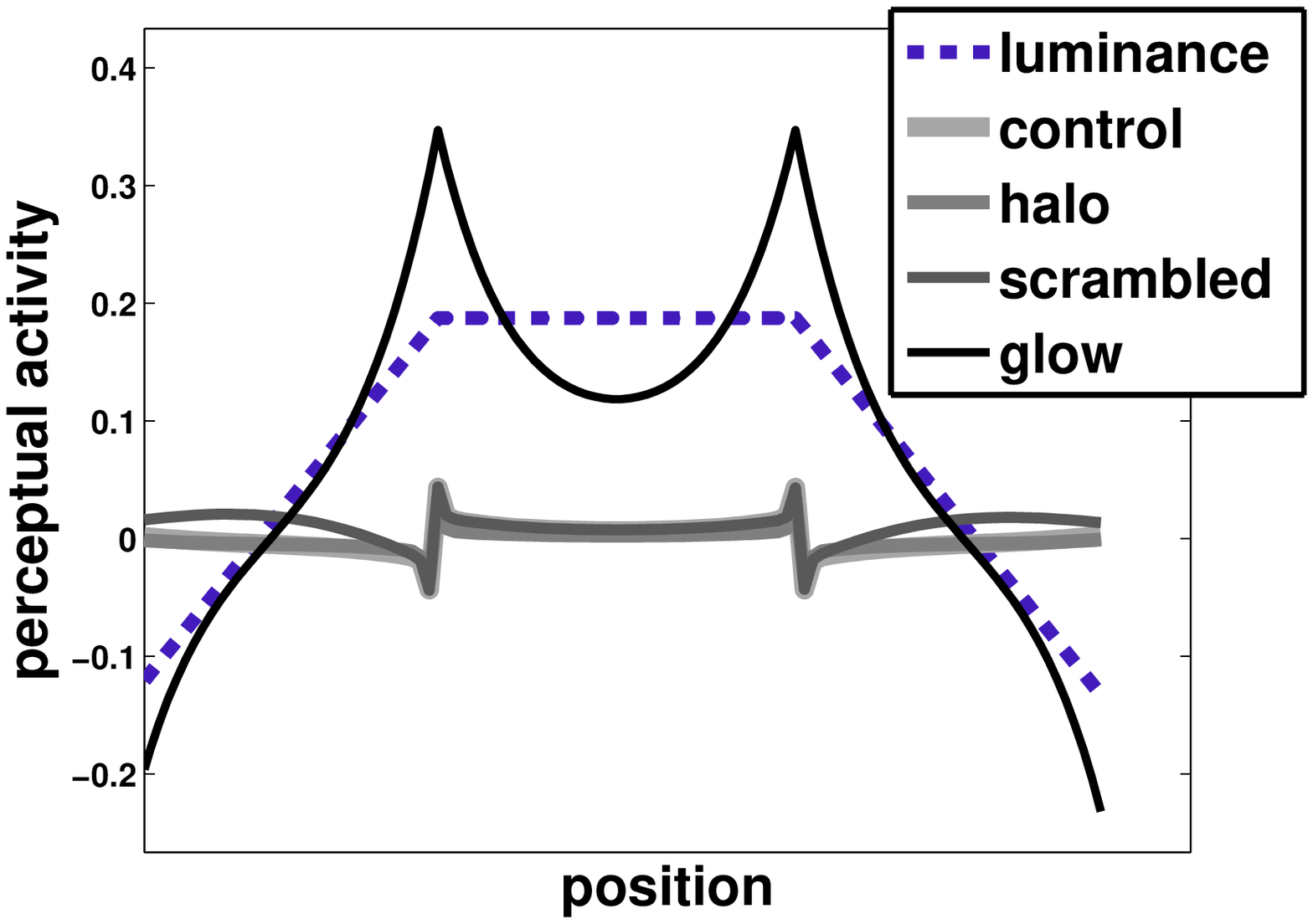}}\\
	  {\bf \large (b)}\scalebox{0.45}{\includegraphics{\pics/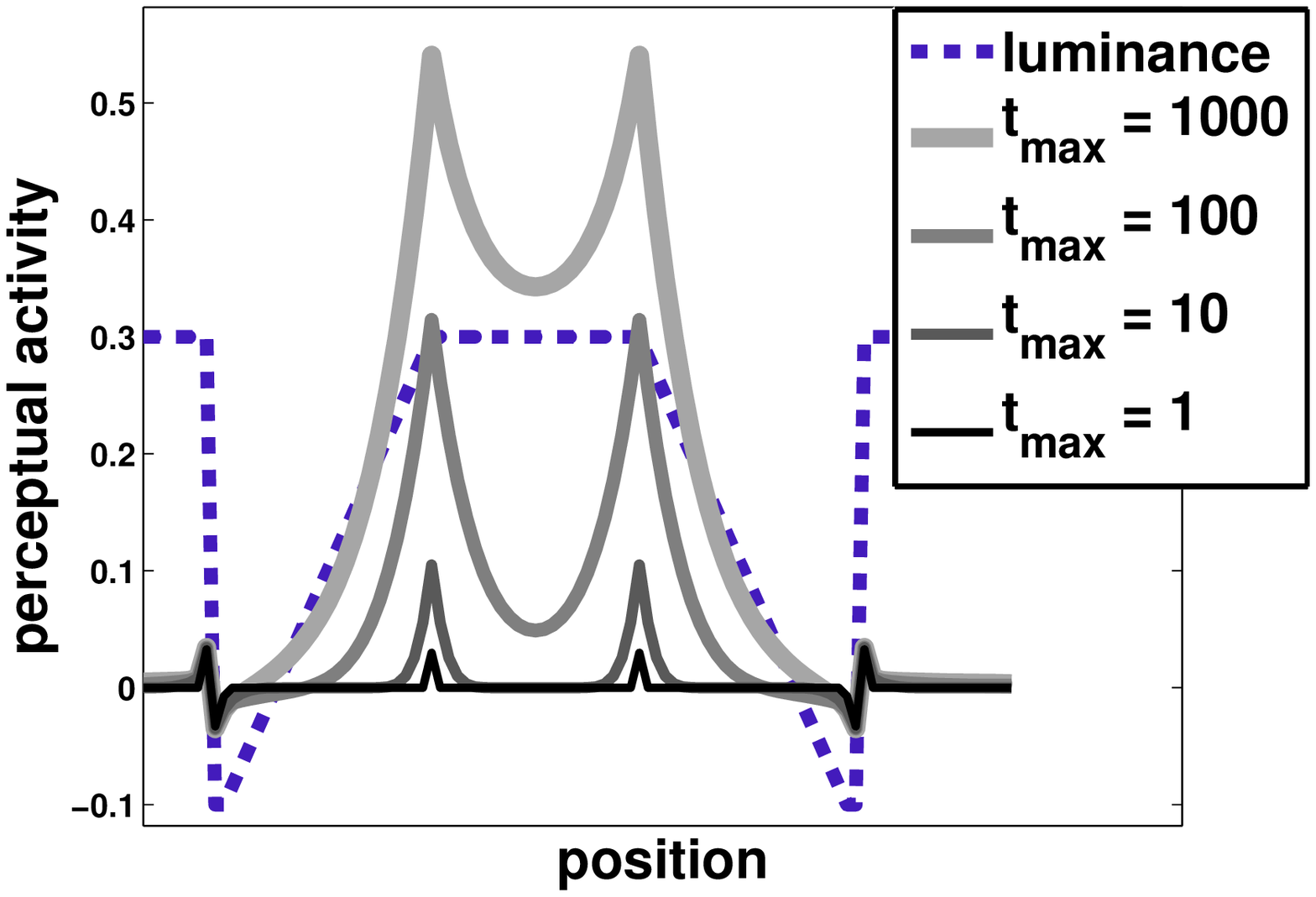}}
	\Caption[GlowProfiles][Profiles of gradient representations shown in \fig[setups]][{%
	{\bf (a)} The curves correspond to horizontal profiles of the 2-D gradient representations
	shown in \fig[setups] (profiles show all columns for the center row of a 2-D display).
	Each curve thus represents a different setup as denoted by the legend.
	{\bf (b)} Profile plots of gradient representations at different simulation times
	(see legend) for the second image of \fig[GlareParade] (spatial frequency $2.5$
	cycles per image).  Note the elevated gradient brightness activity (positive values)
	across the central square.  At time $1$, non-zero gradient activity is obtained
	only at Mach band positions, but not across the central square.  However, during
	the generation of the gradient representation, the central square gets filled in
	with brightness.  This filling-in effect occurs only for the \glow setup, but not
	for any of the others.  Curves representing luminance were rescaled independently
	for both plots (original luminance values ranged always from $0$ to $1$).}]
\end{figure}
%
%
\subsection{Simulations of different setups}
The bottom row of \fig[setups] (``gradients'') shows gradient representations
which have been generated from the images shown in the first row (``setup'').
The gradient representation produced by the \glow setup shows a neon-like
square that is located along the contours of the central square.  Because linear
gradients (i.e., luminance ramps) were used as inducers, each side of the neon-square
actually corresponds to a bright Mach-band \Cite{Mach1865}.  In the course of clamped
diffusion dynamics (second stage of the gradient system), the Mach bands implement brightness
sources, from which activity spreads laterally to generate representations of luminance
ramps (the dark Mach band constitutes the corresponding brightness sink).
In the \glow setup, the four Mach bands are situated around the central square,
thereby forming a closed region where gradient brightness accumulates over time
(\fig[GlowProfiles]\emph{b} \&  \ref{GlowSketch}).  In other words, although
there is no (physical) luminance gradient present across the central square,
it is ``tagged'' with strong gradient brightness.  Because activity does not
dissipate (i.e., there is no drain or brightness sink across the target),
and because brightness sources constantly generate activity, overall brightness
activity eventually grows higher than darkness activity. Thus,
``perceived brightness'' is higher than ``perceived darkness'' in the final
representation\footnote{The gradient systems makes symmetrical predictions
for brightness and darkness, and therefore produces analogous results for
inverse displays (darkness enhancement when the central square is black,
and the ramps terminate with black at the central square).  I put the
terms ``perceived brightness / darkness'' in quotes because brightness
is ``perceived luminance'' and thus already refers to a perceptual variable.},
and the central square will appear luminous.\\
In the \control setup, no luminance gradients are present.
The corresponding gradient representation has low activity, with
similar amplitudes of brightness and darkness.  Due to the
absence of brightness and darkness sources, no lateral spread
of activity occurs, and activity across the central square is close
to zero (as indicated by gray colors in \fig[setups], see also
\fig[GlowProfiles]\emph{a}).
In the \scram setup, again bright Mach bands (i.e., brightness sources)
are created. However, the contour of each ramp, along which luminance
increases, contrasts strongly with the central square.  These contrasts
are ``non-gradients'' and constitute barriers for the propagation of
gradient activity.
Thus, no brightness activity originating from the Mach bands can
propagate into the central square, and no brightness enhancement
of the latter occurs.
The gradient representation that is created for the \halo setup
is similar to the \control setup.  Notice, however, that neither
bright Mach bands nor activity gradients are created at the bright
side of each ramp.  This is due to a strong contrast
with the domain boundary, as a consequence of the domain boundary
conditions which were used for the simulation (c.f. \cite{MatsNC06}
or \cite{gradVisRes06}).\\
The predictions of the gradient system can be summarized as follows
(c.f.\fig[GlowSketch]).
A target region is perceived as being light-emitting if in its gradient
representation it is tagged with high brightness activity, despite of the
absence of actual luminance gradients across that target. 
The target is filled in with brightness if \emph{(i)} brightness
sources are located sufficiently close to it, and if \emph{(ii)} no
activity is annihilated because of the presence of drains or sinks
nearby or across the target.  Then, brightness can accumulate
(i.e., activity grows in time across the target region), and
finally gets much higher than darkness activity, such that
luminosity is perceived (a strong excess of brightness over
darkness).  This situation is typically created by the presence
of linear luminance gradients adjacent to the target.
These predictions are examined further in the following section
by introducing specific modifications of the original \glow
setup.
%
\begin{figure*}[ht!]
	\begin{center}
		\scalebox{0.5}{\includegraphics{\pics/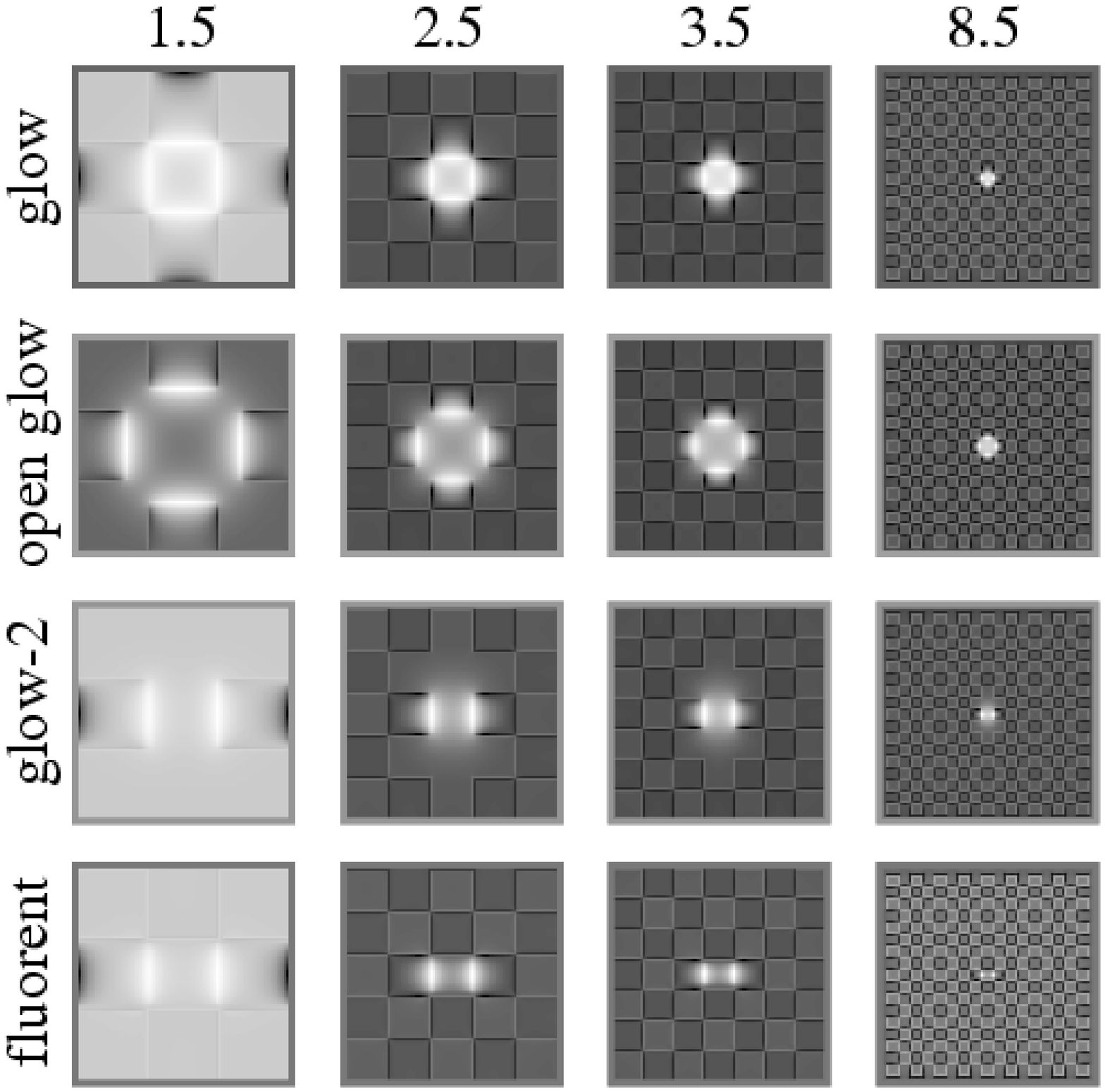}}
	\end{center}
	\Caption[GradsGlowParade][Gradient Representations for \fig[GlareParade]][{%
	Each image has been normalized individually to improve visualization.
	Darkness activity of gradient representations corresponds to dark colors, and
	brightness activity to bright colors.}]
\end{figure*}
\begin{figure}
	 \scalebox{0.45}{\includegraphics{\pics/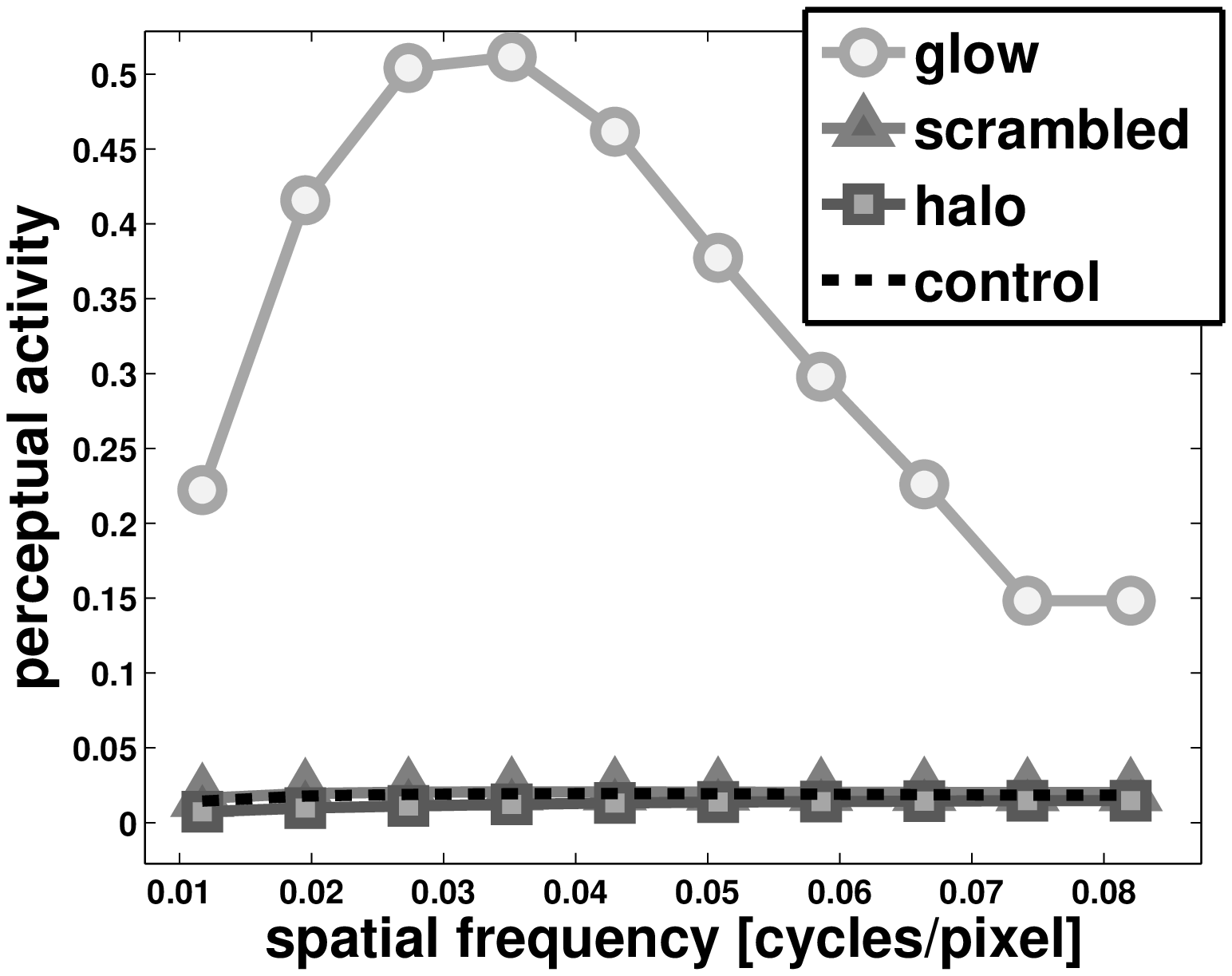}}
	\Caption[ChessK][Varying spatial frequency (setups)][{The curves show gradient
	activity averaged over the central square of the glare effect display
	for various spatial frequencies of the chessboard carrier.  Each curve
	represent a different setup (legend -- see \fig[setups]).  The gradient
	system predicts relative high activities across the central square only
	for the \glow setup, where humans perceive the central square as being luminous.}]
\end{figure}
\begin{figure}
	  {\bf \large (a)}\scalebox{0.45}{\includegraphics{\pics/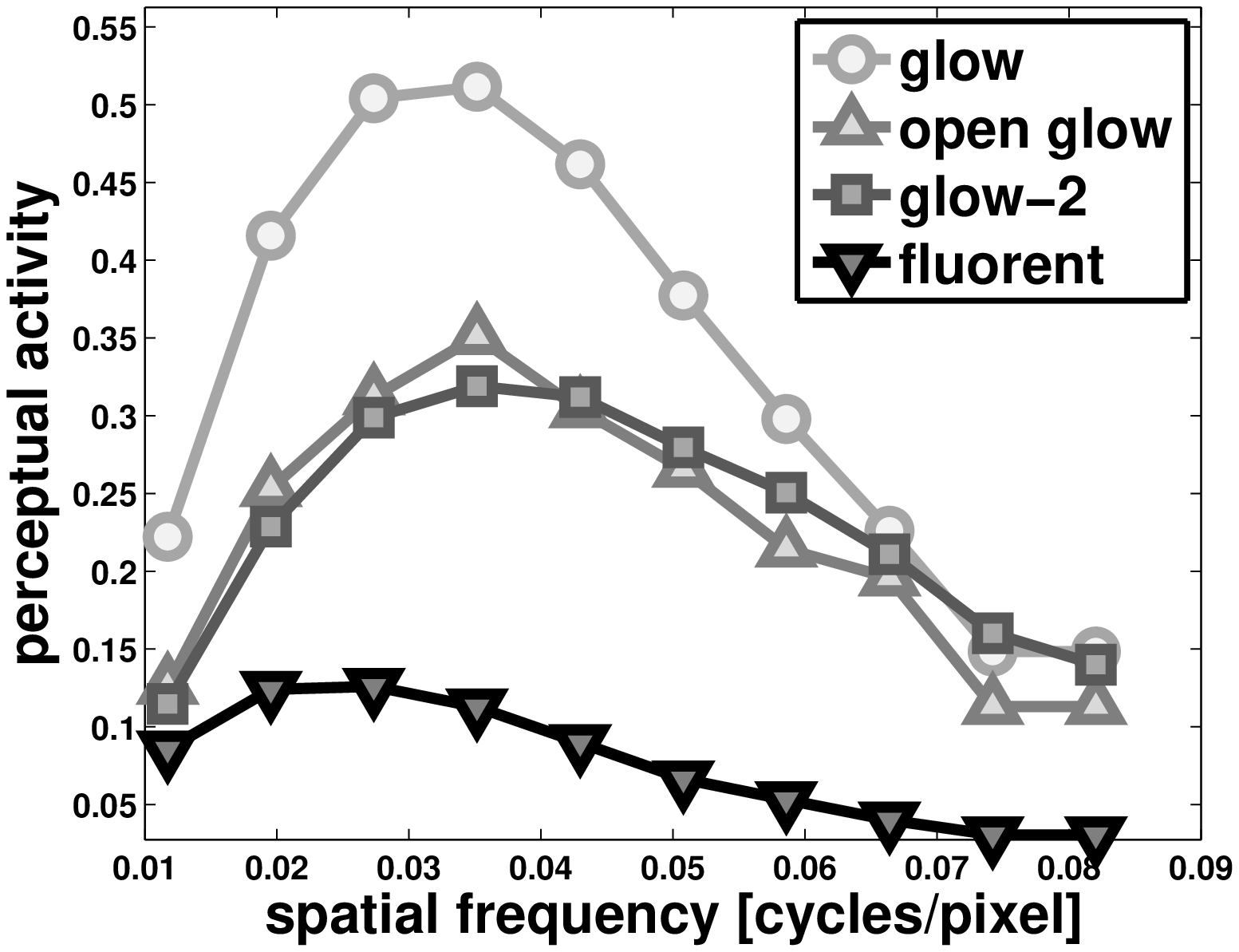}}
	  {\bf \large (b)}\scalebox{0.45}{\includegraphics{\pics/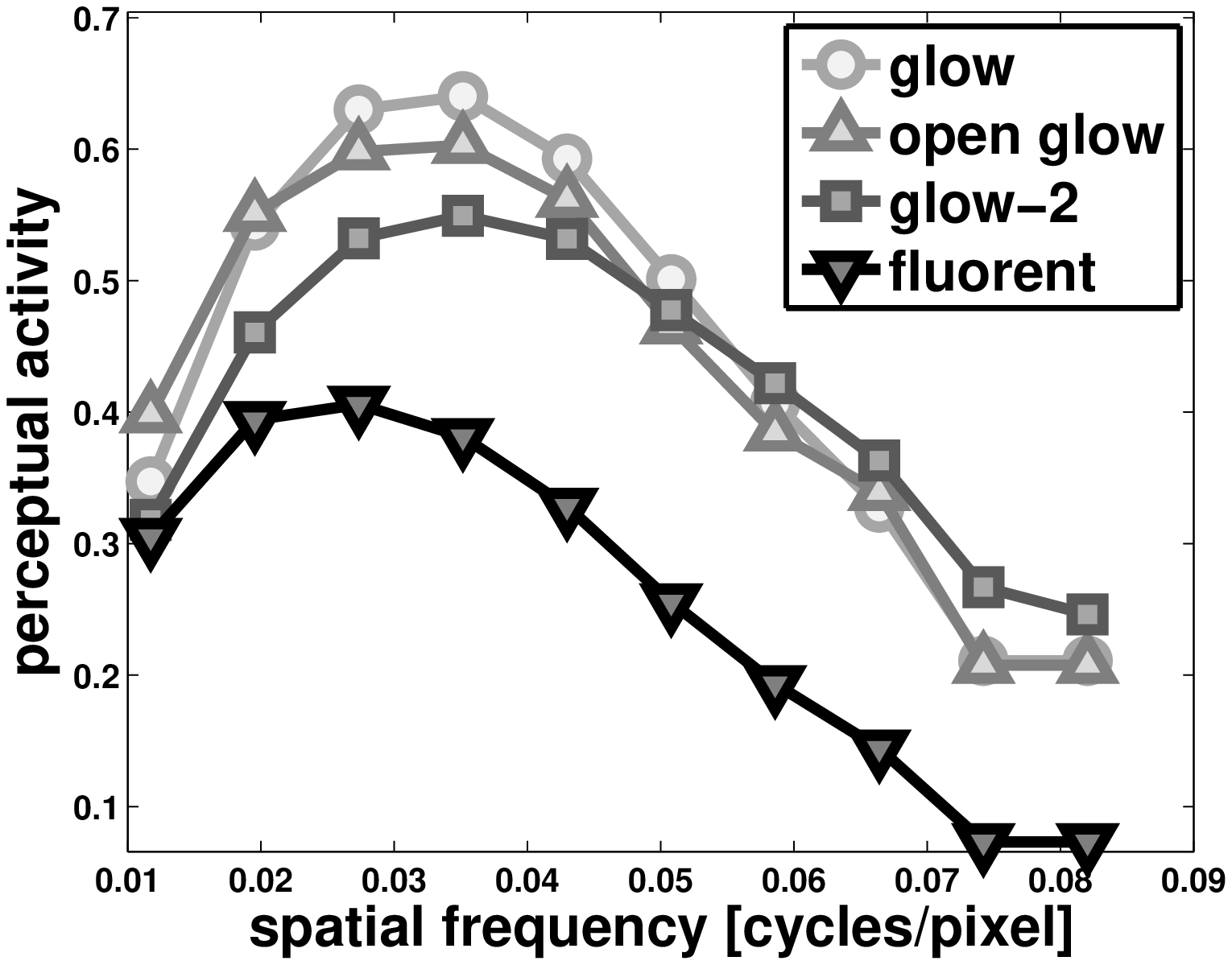}}
	\Caption[displays_ChessK][Varying spatial frequency (displays)][{%
	Both plots show the gradient activity of the central square for the different types
	of luminance displays shown in \fig[GlareParade] with spatial frequencies indicated
	on the abscissa.  Display types are denoted in the legend.  For the open glow display,
	the central square region for measuring gradient activity had to be expanded by $32\%$
	to each side in order to capture the full effect.
	{\bf (a)} The gradient activity is measured by spatial averaging activity values over
	the central square. 
	{\bf (b)} The maximum value of gradient activity over the central square was computed.}]
\end{figure}
%
\subsection{Modifications of the glare effect display and size effects}
\Fig[GlareParade] shows three modification of the original glare effect
display which also lead to the perception of luminosity.  The corresponding
gradient representations are shown in \fig[GradsGlowParade].
\begin{description}
\item[Open glow.] Each luminance ramp was shifted by $32\%$ (of the square length
in pixels) to the darker side,
and the total ramp size was reduced to $75\%$.  Still a glowing effect can be
observed.  The gradient system consistently predicts this effect -- brightness
of the ramp accumulates in the central part of the image, although activity
propagation now takes place over a larger region than the central region
of the original display, and despite of the target region being no longer
tightly enclosed by the Mach bands.
\item[Glow-2.] A glow effect is also seen with only two luminance ramps.  However,
this effect is weaker because brightness activity can escape at the top and the
bottom into the white regions adjacent to the target region.
\item[Fluorent.] The top and the bottom side of the central square is now enclosed
by uniform black squares.  The boundaries of each black square give rise to non-gradient
inhibition, thus implementing activity drains at the central square (\fig[FluorentSketch]).
Therefore, brightness enhancement should be weaker compared to the \emph{glow-2} display.
In fact, the sensation appears to be what has been described as ``preluminous
super white'' \Cite{Heinemann55} or ``\emph{fluorent}'' \Cite{Evans1959}.

\end{description}
%
%
\begin{figure}
	\scalebox{0.4}{\includegraphics{\pics/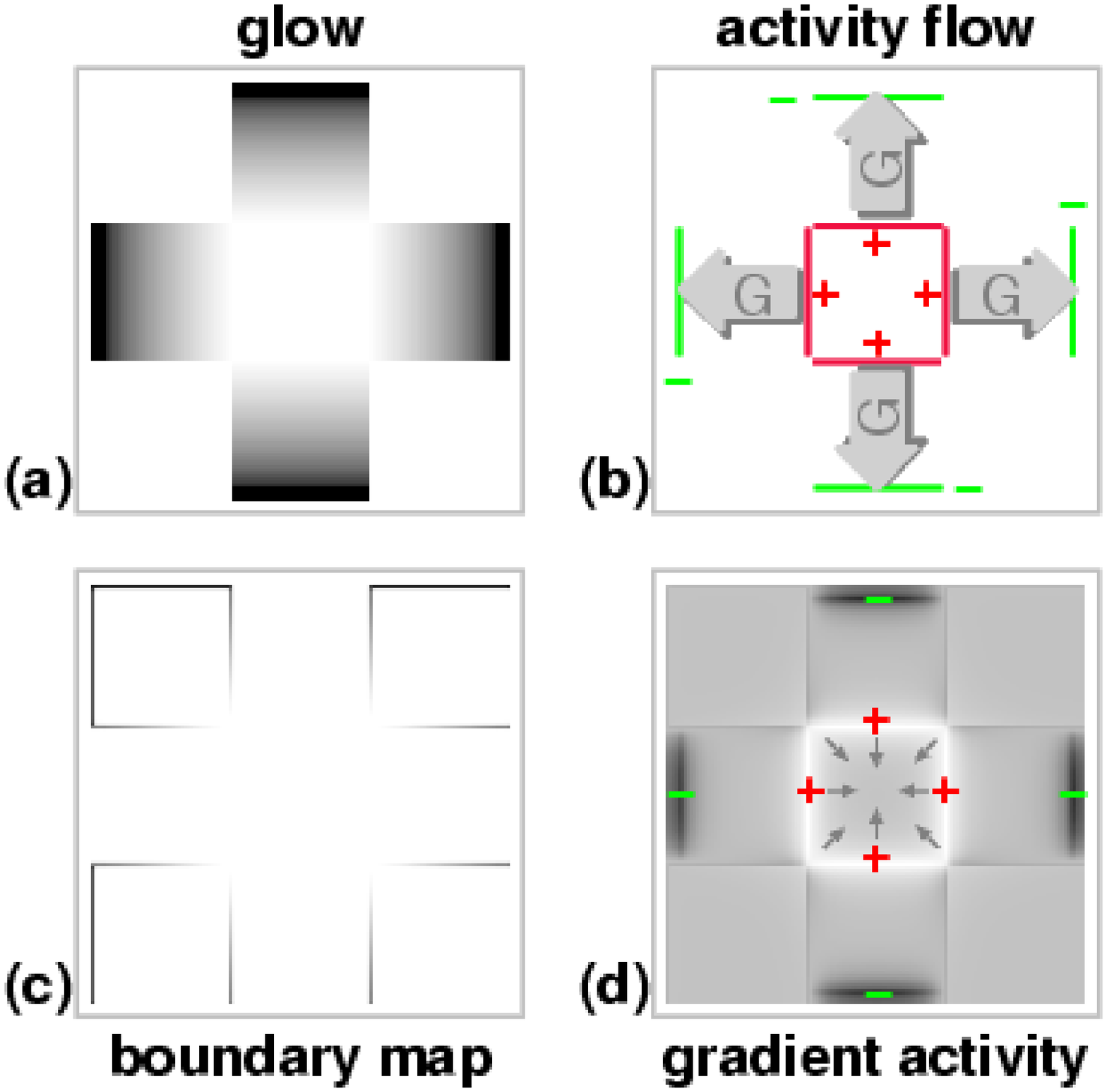}}
	\Caption[GlowSketch][Why the glow display produces the strongest effect][{%
	{\bf (a)} The glow display (c.f. \fig[GlareParade])
	{\bf (b)} (Brightness) sources are designated with ``$+$'', and (brightness)
	sinks with ``$-$''.  An activity gradient will form between sources and sinks:
	The arrows designated by ``G'' indicate the direction of gradient formation
	from increasing to decreasing perceived luminance.  Notice that the central
	(or target) square is surrounded by four brightness sources.
	{\bf (c)} The boundary map is equivalent to locations where non-gradient
	inhibition is active.  It is assumed that surface representations are
	triggered there.  No boundaries are present around the target square, and
	thus no non-gradient inhibition is produced.
	{\bf (d)} Since brightness sources constantly generate activity, and
	no loss of activity occurs across the central square (due to the
	presence of a brightness sink or non-gradient inhibition), brightness
	activity can accumulate (small arrows; see also
	\fig[GlowProfiles]).  Accumulated brightness over the target square
	is proposed to be associated with the perception of luminosity.
	Notice that the activity gradients do not extend into the four
	white squares in the corners because of non-gradient inhibition.}]
\end{figure}
\begin{figure}
	\scalebox{0.4}{\includegraphics{\pics/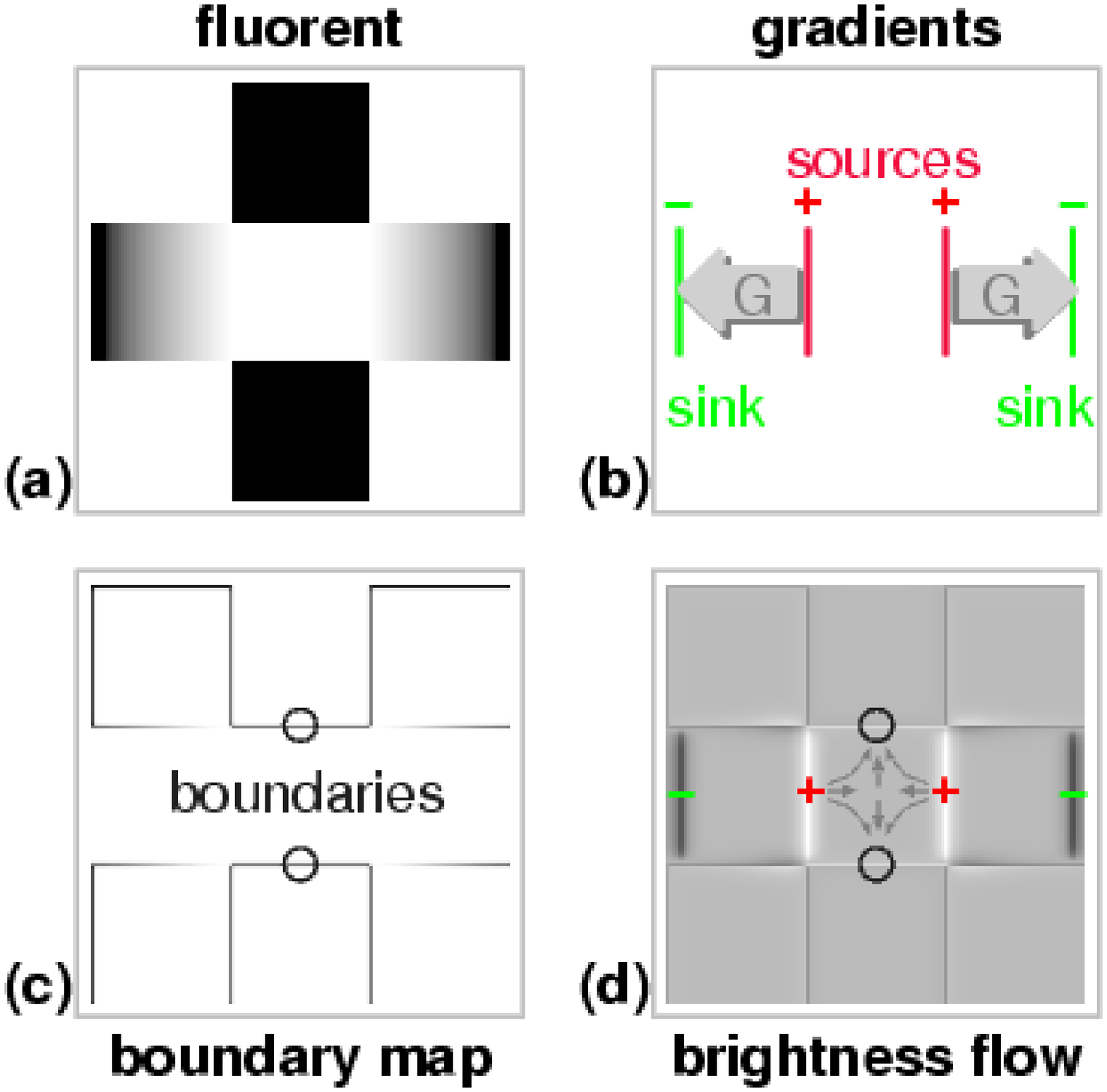}}
	\Caption[FluorentSketch][Why the fluorent display produces the weakest effect][{%
 	{\bf (a)} The fluorent display (c.f. \fig[GlareParade])
	{\bf (b)} In comparison to the glow display (previous figure), only two
	activity gradients will be generated (arrows designated with ``G'').
	{\bf (c)} Apart from two brightness sources, the central square is now
	also flanked by two contours (``$\bigcirc$'') giving rise to non-gradient
	inhibition.  As explained in the last paragraph of section \ref{subsect:howitworks},
	non-gradient inhibition acts like a passive drain, for both brightness sources and
	brightness sinks.
	{\bf (d)} Activity propagates from brightness sources ``$+$'' into the central square,
	but its accumulation is less than with the glow display (\fig[GlowSketch]) because
	it gets annihilated at contours ``$\bigcirc$'' (small arrows).  Therefore, the target
	square of the fluorent display will appear less luminous than the target of the
	glow display.}]
\end{figure}
Because all of the glow effects presented in this paper are induced by linear luminance
ramps, and because Mach bands are attached to linear luminance ramps, the glow effects
should also depend on ramp width or scale, respectively.  The perceived
strength of Mach bands is small for narrow ramps, large at ramps of intermediate size,
and decreases again with broad ramps (``inverted-U''-behavior, \cite{RossMorroneBurr89}). 
Increasing the spatial frequency of the chessboard carrier decreases both the ramp
width, and the size of the central square.  \Fig[GlareParade]
illustrates the dependence of the perceived glowing strength on spatial frequency
for $1.5$, $2.5$, $3.5$, and $8.5$ cycles per image.  Although precise psychophysical
data concerning this spatial frequency dependence are not (yet) available, some of
the effects seem to be stronger at an intermediate frequency.  The gradient system
clearly suggests a relationship between carrier frequency and
glow strength (\fig[GradsGlowParade]).  Notice, however, that the gradient system
is not calibrated with respect to viewing distance, and maximum effects may be
predicted at different spatial frequencies than perceived by humans when looking
at \fig[GlareParade].\\
In \fig[ChessK], the strength of glowing is quantified in terms of the mean gradient
activity over the central square for different spatial frequencies of the chessboard
carrier (see section \ref{sect:methods}).  A maximum effect is predicted for the
\glow setup, but no brightness enhancement of the target does occur for the setups
\scramx, \halox, and \controlx.
In \fig[displays_ChessK], the strength of glowing is measured both by computing
the mean activity over the central square (\emph{a}) and the maximum (\emph{b}).
The glow display is predicted to produce the strongest effect (\fig[GlowSketch]),
and the fluorent display to produce the weakest (\fig[FluorentSketch]).
The important result with these curves
is the prediction of a maximum at some intermediate spatial frequency.  Notice,
however, that the curves shift along the ordinate depending on whether the spatial
average across the central square was computed, or the maximum value was taken.
This is because the central square is not filled in homogeneously with brightness
activity, but gradient activity rather decreases towards the center of the
central square (\fig[GlowProfiles]).  This ``bowing effect'' is especially
prominent with larger region sizes or at low spatial frequencies, respectively
(cf. \fig[GradsGlowParade]).
%
\begin{figure*}
	  \scalebox{0.45}{\includegraphics{\pics/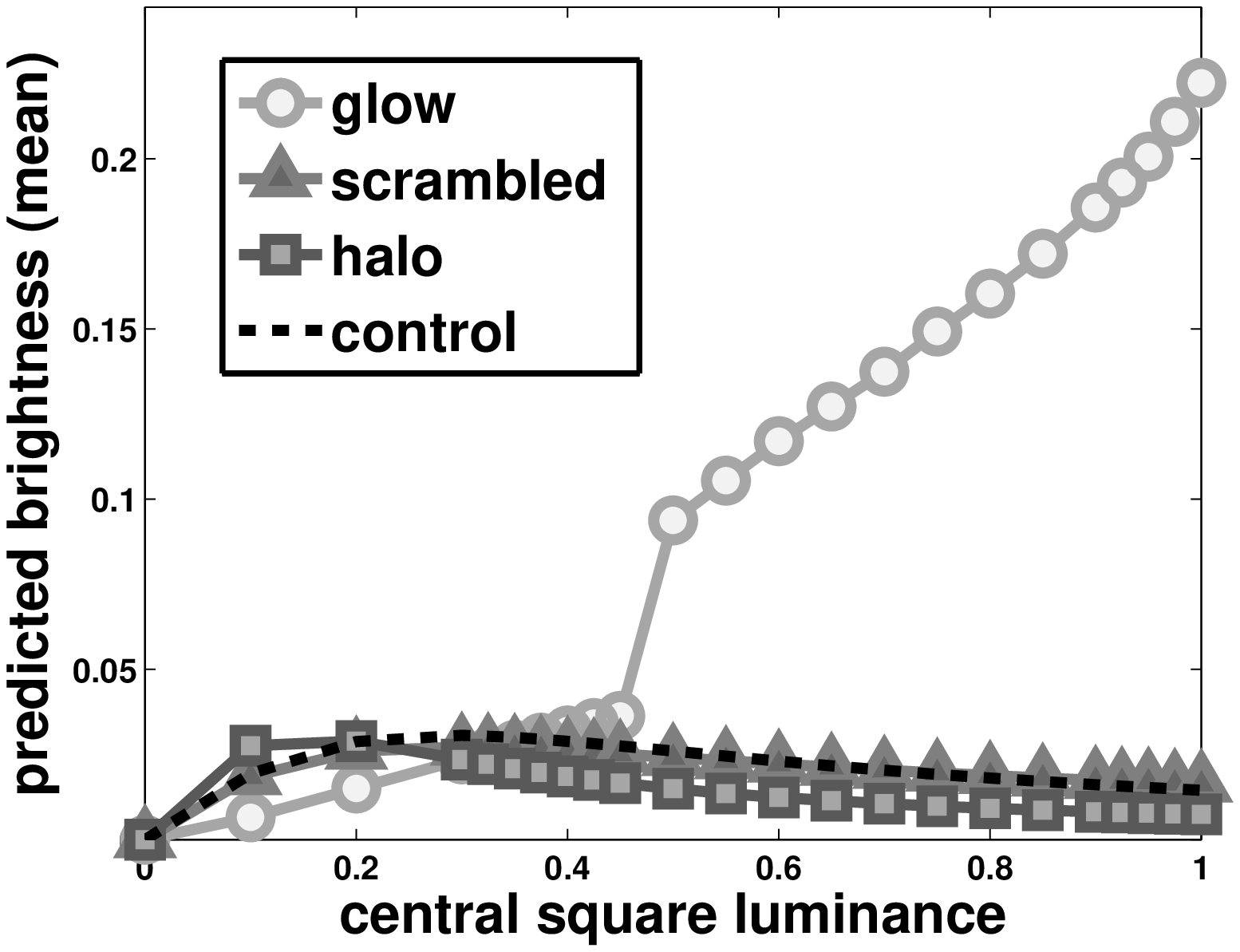}}
	  \scalebox{0.75}{\includegraphics{\pics/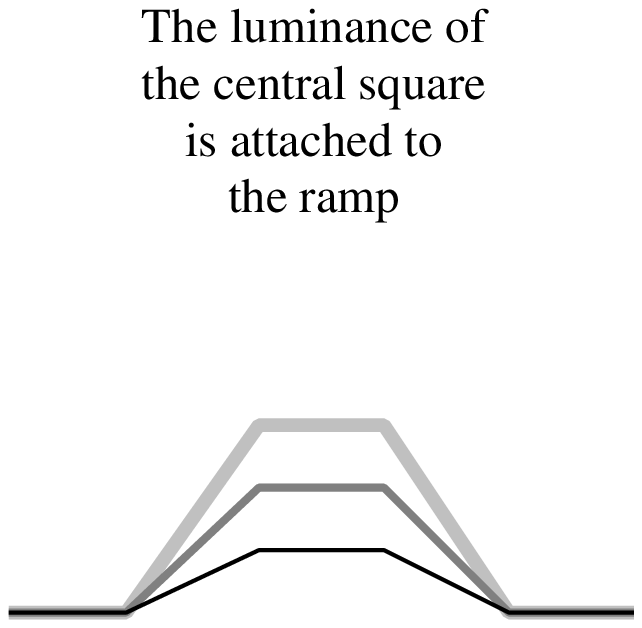}}
	\Caption[CentralSquareVaries][Varying the intensity of the central square][{%
	The curves show the predicted gradient brightness activity as a function of
	the luminance of the central square (sketch) for the setups of the glare effect
	display ($1.5$ cycles per image, cf. first image in \fig[GlareParade]).  The
	gradient brightness associated with the \glow setup abruptly increases
	between intensity levels $0.4$ and $0.5$ ($=$ step-like increment),
	whereas a relatively weak dependence on luminance is predicted for the
	setups \scramx, \halox, and \controlx.  Similar curves are obtained by
	plotting the maximum activity of the central square.  The location of
	the step-like increment does not depend significantly on the spatial
	frequency of the chessboard carrier.}]
\end{figure*}
%
\subsection{Glowing grays?}
\cite{ZavagnoCaputo2005} reported the perception of ``glowing grays''
\Cite{Wallach48} in a psychophysical experiment where subjects first
had to adjust the central square of a chessboard display until it was
perceived as white (no luminance gradient was present in this display).
Next, they were asked to adjust the central square of a second display
until it was perceived to glow (the image for their second display was
identical to the \glow setup in \fig[setups]).  The experiment was carried
out for three different luminance levels of the (originally white) squares
in the corners of the display ($=$ background luminance).  The authors observed
that subjects did not adjust the central square of the \glow display to white.
In other words, it was already perceived as glowing at some gray level.\\
\Fig[CentralSquareVaries] shows the dependence of gradient brightness
on the luminance level of the central square.  Notice that the curve
for the the \glow setup reveals an abrupt increase between luminance
levels $0.4$ and $0.5$.  In other words, the gradient system reveals
a threshold behavior\footnote{I verified that the step indeed corresponds
to a threshold by repeating the simulations by choosing smaller increments
of luminance.  The step always appeared no matter how small the luminance
increments were chosen.}, where gradient brightness strongly increases
with the respect to the curves for the other setups \scramx, \halox, and
\controlx .  After the step-like increment, the curve shows an approximately
linear dependence on the luminance level of the central square.  At luminance
$\approx 0.9$, the curve reveals a moderate increase in slope.\\
Because only the \glow setup leads to the sensation of glow,
and because before the step-like increment gradient brightness
is approximately the same as with the other three setups (which are not
associated with the perception of glow), this step-like increment
in fact corresponds to an absolute threshold for the central
square to be perceived as light-emitting.  
Moreover, because the step-like increment occurs between luminance
levels $0.4$ and $0.5$ which is associated with mid-gray, the gradient
system indeed predicts the occurrence of ``glowing grays''.\\
The background luminance level influences in retinal adaptation, and
also in lightness anchoring.  The present version of the gradient
system, however, does neither incorporate mechanisms for adaptation,
nor does it incorporate interactions with surface representations
(or lightness computations).
%
\begin{figure*}
	  \scalebox{0.45}{\includegraphics{\pics/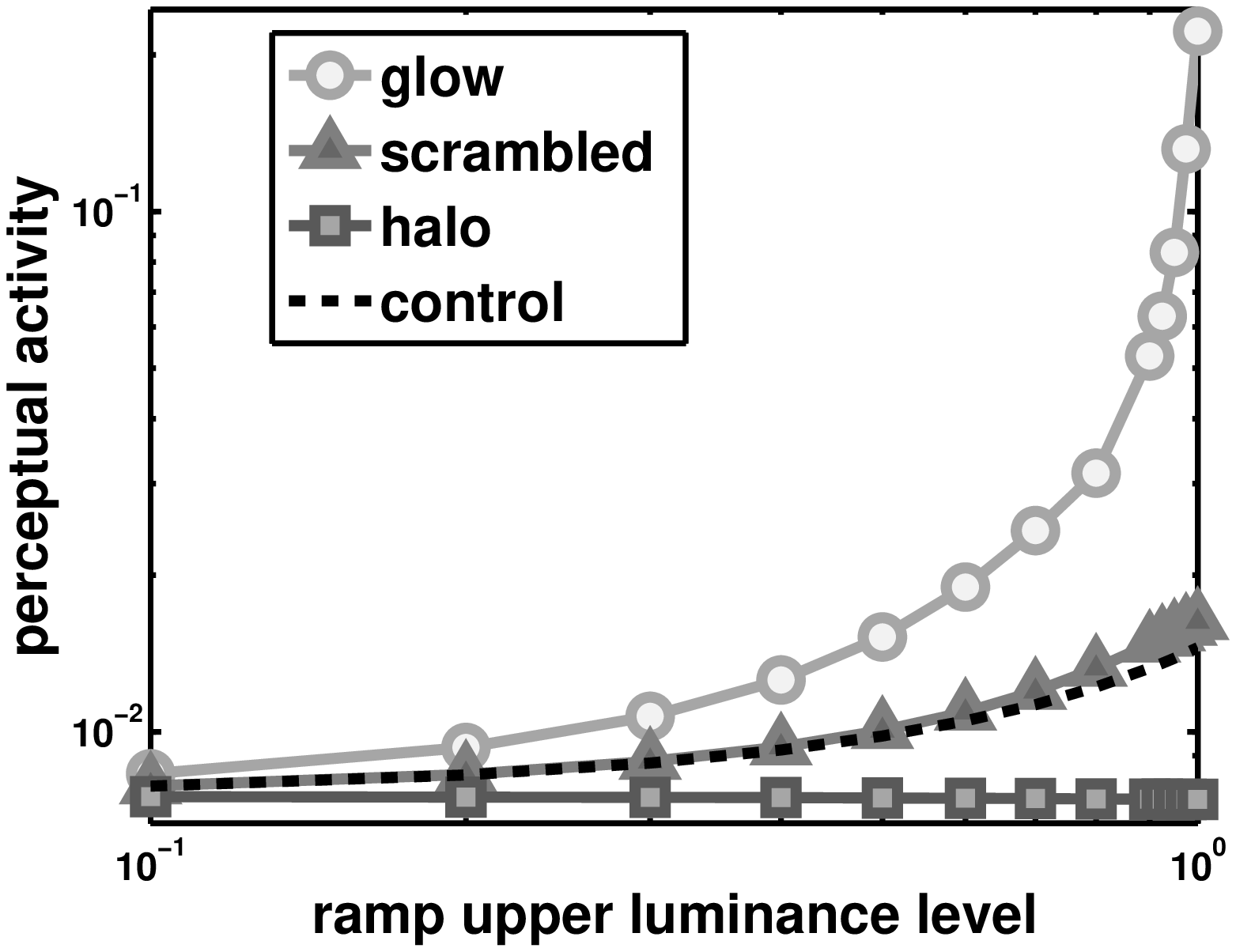}}
	  \scalebox{0.75}{\includegraphics{\pics/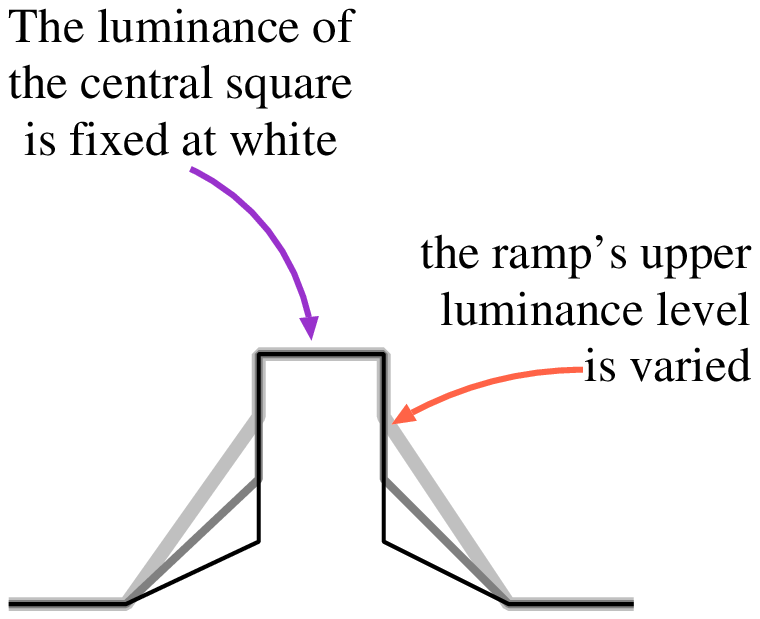}}
	\Caption[CentralSquareFixed][Varying intensity][{%
	The luminance of the central square of the \chessramp\s display was held fixed at white,
	and the upper knee-point of the ramp was set to different luminance values from black
	(corresponding to an ordinary chessboard without luminance ramps) to white (corresponding
	to the glare effect display).  The shape of the curve for the \glow setup matches well
	the psychophysically measured curve shown in figure 3 of \cite{ZavagnoCaputo2001} (but
	see text for further details).}]
\end{figure*}
%
%
\subsection{Influence of the luminance ramp}
In \cite{ZavagnoCaputo2001}, subjects were asked to adjust the height of the luminance ramps
surrounding the central square of a glare effect display ($=$ first image of \fig[setups])
while the central square was always held fixed at white.  This procedure was
repeated for different levels of the four (originally white) squares in the corner of the display
($=$ background luminance).  The authors found that the threshold for perceiving the central
square as glowing ($=$ luminosity threshold) increased with increasing background luminance
in precisely the same way as the curve for the \glow setup in \fig[CentralSquareFixed]
(see figure 3 in \cite{ZavagnoCaputo2001}).  However, \fig[CentralSquareFixed] does not
show luminosity threshold versus background luminance, but gradient activity of the
central square versus the upper ramp luminance.  So why is it that both curves are
so similar?\\
The curve of \fig[CentralSquareFixed] (\glow setup) shows the predicted sensation
of luminosity given some ramp luminance level (as illustrated by figure 1 in
\cite{ZavagnoCaputo2005}).  The results from \cite{ZavagnoCaputo2001}, demonstrate
that the luminosity threshold increases as a function of background luminance.
When comparing their results to the predictions of the gradient system, it
therefore seems that the background luminance level sets a baseline level
below of which luminosity cannot be perceived.  This idea is equivalent to
putting horizontal lines in \fig[CentralSquareFixed], with an intercept
proportional to background luminance.  Therefore, to perceive luminosity,
the upper ramp luminance has to be adjusted such that gradient activity
is just above the horizontal line.  And this is what is shown in figure
1 of \cite{ZavagnoCaputo2001}.
%
\section{Discussion}
\label{sect:discussion}
%
A recent psychophysical study from \cite{CorreaniSamuelLeonards06}, assigned feature status
to luminance gradients.  Accordingly, here I studied the predictions of the gradient
system for luminance displays which appear self-luminous.  The gradient system is an
instantiation of a recently proposed theory about how luminance gradients are segregated
from images, and how representations of luminance gradients are generated
\Cite{MatsNC06,gradVisRes06}.\\
Here I showed that gradient representations have higher activity levels across the central
square in those displays
that are associated with the sensation of glow (\glow setup, \fig[setups]).  Conversely,
gradient activity is low in displays which are not perceived as light-emitting (setups \controlx,
\halox, \scramx).  My results therefore support the conjecture from \cite{ZavagnoCaputo2005}
that luminance gradients play a crucial role in luminosity perception.\\
Three modifications of the glare effect display were devised (\fig[GlareParade])
to put to the test the following predictions of the gradient system (\fig[GradsGlowParade]):
\emph{(i)} gradient activity accumulates in the central square what predicts a corresponding
enhancement in perceived brightness (original ``glow'' display);
\emph{(ii)} if the central square is enclosed by only two luminance ramps (``glow-2'' display),
then gradient activity spreads into the open region, but luminosity should still be perceived;
\emph{(iii)} if the central square is delineated by sharp contrasts (``fluorent'' display),
then drains for brightness activities are created which should lead to a reduction of the
glow effect, and finally
\emph{(iv)} as gradient activity spreads laterally originating from brightness sources
(which are perceived as bright Mach bands surrounding the central square), the perceived
luminosity should depend on the size of the target region or the spatial frequency of the
chessboard carrier, respectively (``open glow'' display, and \fig[ChessK] \& \fig[displays_ChessK]).\\
For the luminance displays considered in this paper, the gradient system predicts
a threshold behavior above which the central square is perceived as being
light-emitting (the step-like change in \fig[CentralSquareVaries]): Light emission
is already predicted at intermediate luminance values (``glowing grays'').  In addition,
the gradient system provides a consistent explanation of the results from
\cite{ZavagnoCaputo2001} (compare their figure 1 with my \fig[CentralSquareFixed]).\\
%
\begin{figure}[ht!]
	\begin{center}
		\scalebox{0.35}{\includegraphics{\pics/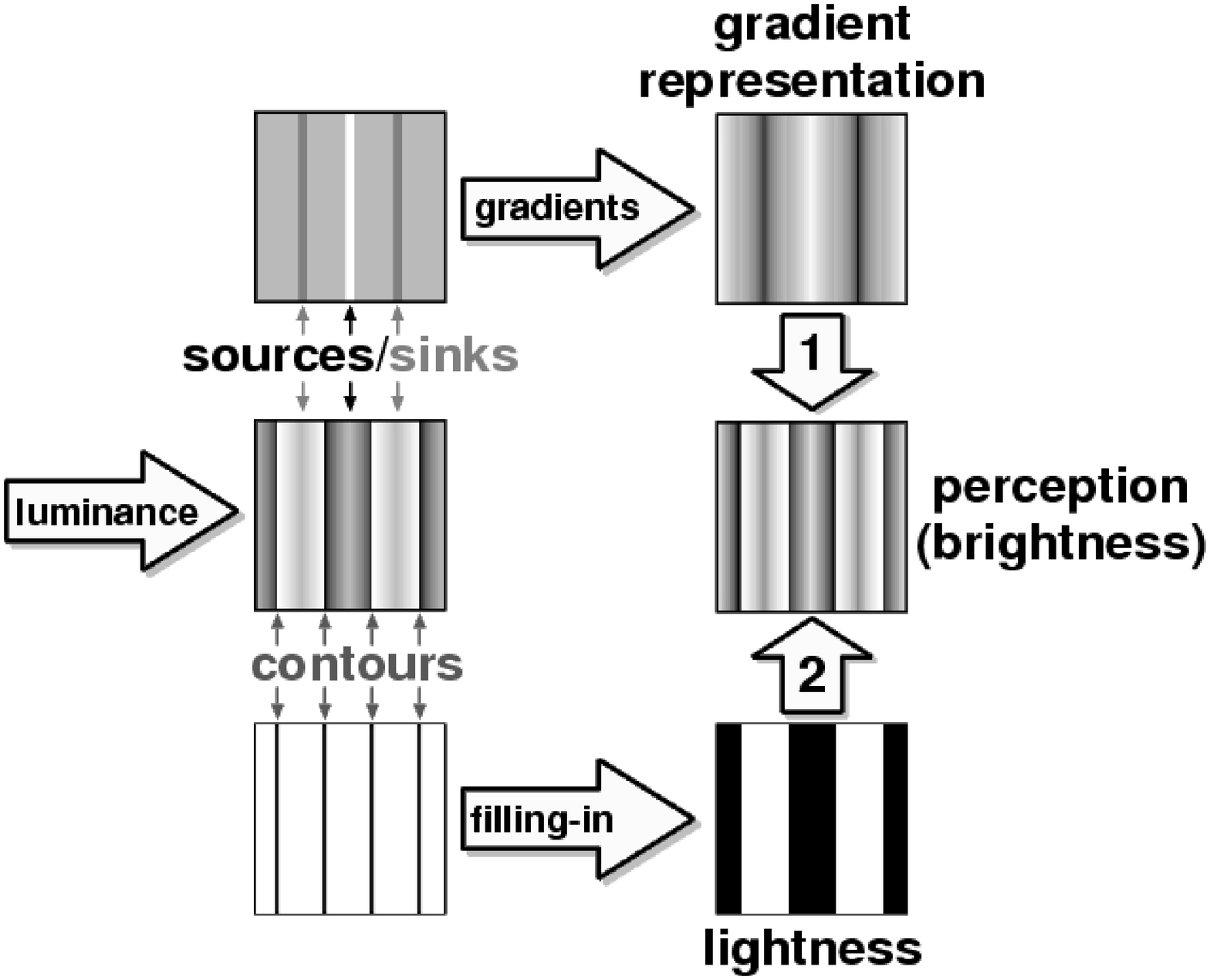}}
	\end{center}
	\Caption[interactions][How gradient representations relate to the filling-in framework][{%
	A ``notch grating'' luminance pattern (``luminance'' arrow) is used to illustrate
	conceivable interactions between lightness computations and gradient representations.
	Contours are detected at the step-like luminance changes, what triggers a filling-in
	process for computing surface lightness (``filling-in'' arrow; \cite{GerritsVendrik70,CohenGrossberg84}).
	Filling-in processes were suggested as a theoretical mechanism to implement invariance properties
	for surface representations, for example``discounting the illuminant'' to implement lightness constancy
	(e.g., \cite{GrossTodo88,PeMiNe95,PesNeu98,GroPes98,NeumannEtAl01}).
	As we perceive lightness constancy, but at the same time also smooth changes in luminance
	\Cite{ToddFarleyMingolla04}, surface lightness and gradient representations need to
	interact (arrows ``1'' \& ``2'').  This interaction finally is proposed to result in
	brightness perception.}]
\end{figure}
\begin{figure}
	\begin{center}
		\scalebox{0.45}{\includegraphics{\pics/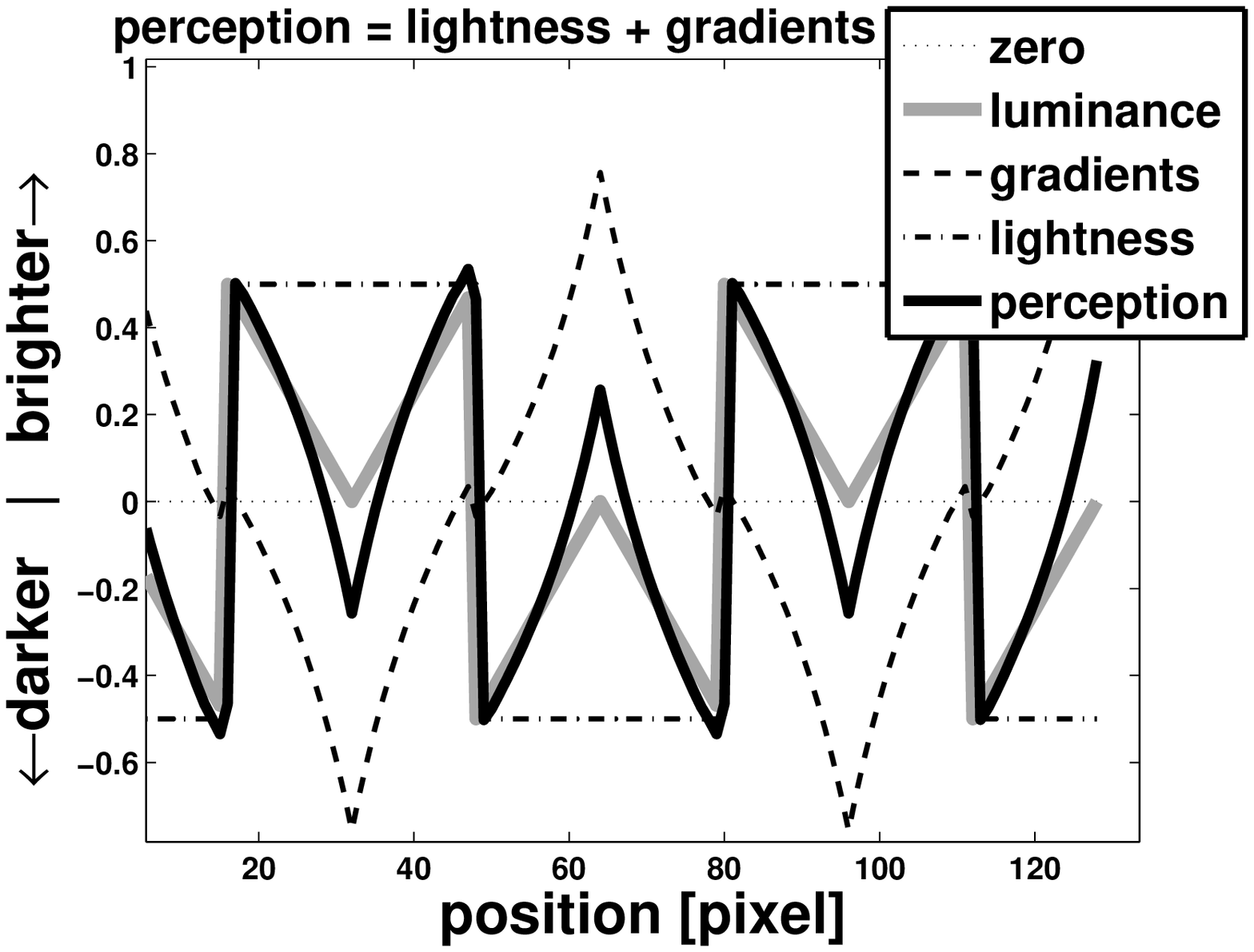}}
	\end{center}
	\Caption[interactions_profiles][Gradient and surface representations][{%
	 The curves illustrate a possible mechanism for the interaction of
	 gradient representations with surface representations.  Each curve
	 shows a profile plot (all columns for a fixed row) of the images
	 shown in \fig[interactions].  The brightness prediction was computed
	 by assuming a simple linear interaction between ON- and OFF-channels
	 of surface and gradients (figure title), yet it correctly predicts
	 that the faint lines in the center of the dark step are perceived
	 brighter than the faint lines of the bright step, albeit both lines
	 have the same luminance level.}]
\end{figure}
%
\subsection{Gradient representations, lightness, and brightness}
%
The gradient system was proposed as one part of three, in parallel acting
processing streams, for generating texture representations\footnote{``Texture''
in this context means fine-grained and even-symmetric contrast features.} and
surface representations (see \cite{MatsThesis,MatsEtAl05}).  Although it is
clear that surface representations and gradient representations have to interact
at some level in the object recognition hierarchy (e.g., in order to derive shape
from shading), it is not clear how such interactions could be
implemented at an early level in the visual system.The original idea was that
whenever odd-symmetrical and sharply bounded contrasts are present in an image,
the corresponding information triggers the generation of surface representations
by a filling-in process.  By contrast, the presence of blur or soft contrasts
trigger representations of luminance gradients (\fig[interactions]
and \ref{interactions_profiles}).\\
The present study suggests that the perception of luminosity is associated
with gradient representations, but not with surface representations.  But
then, surface representations can be directly related to perceived reflectance.
Otherwise expressed, the perceptual correlate of surface representations is lightness.
Reflectance describes a property of surfaces which has the value zero if
the surface absorbs all light (and thus appears black), and the value one
if the surface reflects all light.  Gray levels are represented by intermediate
reflectance values.  Ideally, lightness should follow reflectance.
However, if luminosity effects were explained in terms of reflectance,
this would imply that reflectance values were bigger than one, because
the surface would emit more light than it actually could reflect.\\
Furthermore, lightness constancy implies that reflectance is perceived as
approximately constant despite of variations in illumination conditions.
Lightness constancy seems also not to be affected significantly by the
presence of specular highlights on surfaces \Cite{ToddFarleyMingolla04}.
Gradient representations therefore are supposed to contain all surface
information that otherwise would affect lightness constancy and thus
object recognition.  Taken together, luminosity is not perceived on
the lightness scale, but on the brightness scale.  Brightness comprises
all perceptual aspects of a scene, including lightness and luminosity.
What happens, however, if surface representations and gradient
representations are simultaneously triggered for one and the same region? 
This situation occurs, for example, with a luminance staircase giving rise
to Chevreul's illusion (e.g., figure 8 in \cite{MatsNC06}).  Because
Chevreul's illusion commonly is not described as being light emitting,
it seems that the presence of surface representations weakens the
perceptual impact of gradient representations in terms of brightness.\\
Similarly, for the notch grating stimulus, surface representations and
gradient representations are simultaneously triggered (\fig[interactions]).
More precisely, each step-like change in luminance triggers a surface
representation, and the faint lines centered at each step constitute
the sources and sinks for producing gradient representations
(see arrows with corresponding labels in \fig[interactions]).\\
A possible mechanism for creating surface representations are
so-called filling-in processes \Cite{GerritsVendrik70,CohenGrossberg84},
which serve to assign perceptual properties (e.g., lightness, color, or depth)
to object surfaces (e.g., \cite{GrossTodo88,PeMiNe95,PesNeu98,GroPes98,NeumannEtAl01}).
As a result, surfaces are rendered invariant against smooth changes in the
corresponding stimulus attribute.  For example, smooth luminance gradients
are discounted in surface representations, and in this way lightness constancy
is achieved.  Nevertheless we still perceive luminance gradients, what
leads to the question how gradient representations interact with surface
representations.\\
Consider the \scram setup (\fig[setups]).  Although Mach bands
are perceived, no brightness enhancement of the four white
squares in the corners of the display seems to take place.
For each of these squares, a gradient representation is triggered at
one side (where the Mach bands are observed), and a surface
representation is triggered at the other side (at the lower
or dark end of the luminance ramp).\\
Similarly, the configuration that has been examined with
\fig[CentralSquareFixed] corresponds to a situation where
a sharply-bounded contrast concurs with a ramp (see also
figure 1 in  \cite{ZavagnoCaputo2005}).
But then again, a surface representation (giving rise to
lightness activity) and a gradient representation are
generated at the same time for the central square.
Depending on whether the one or the other representation
has higher activity, one perceives luminosity (only gradient
activity), super white (gradient activity and lightness activity),
or only ordinary ``white'' (i.e. only lightness activity).\\
Taken together, these observations support two conclusions.
First, the mechanism for triggering surface or gradient representations
does not operate in a ``all-or-none'' fashion, but rather operates
continuously.  Second, the concurrent presence of a surface
representation and a gradient representation for a region may reduce
or even abolish the perception of luminosity.  Notice that both
conclusions are not mutually exclusive.\\
%
\subsection{Competing models for luminosity perception}
%
I briefly discuss three different models in turn which could in principle account
for the perception of luminosity.\\
\cite{Ullman1976} suggested an extension to the Retinex theory \Cite{LandMcCann1971,LandRetinex77}
such that light sources can be detected in achromatic Mondrian displays.
The idea is to compute the gradient ratio and the intensity ratio between
adjacent surfaces.  If the ratios are different, then one of the areas
is a light source (see \cite{ZavagnoCaputo2001} for a more detailed
discussion of this model with respect to the glare effect display).
Ullman's model thus links luminosity to lightness computations.\\
By measuring the intensities of surfaces, \cite{BonatoGilchrist1994},
and \cite{BonatoGilchrist1999}, could establish a relationship between
surface area and the luminance value at which the surface appeared as
being luminous ($=$ luminosity threshold).  They found that \emph{(i)}
a 17-fold increase in the surface area lead to a 3-fold increase in the
luminosity threshold, and \emph{(ii)} for a surface to be perceived as
light-emitting, its intensity must be $\approx 1.7$ times larger than 
the intensity of a non-luminous, white surface (under identical illumination
conditions).  To illustrate, consider a simple display
where a surface is divided into a dark region and a lighter region.
The luminance ratio of both areas is held constant.  Let the dark
region initially be small, and now gradually increase its size
with respect to the lighter region.  In this case, the lighter
surface is anchored at white according to the
``\emph{Highest-Luminance-As-White}'' (HLAW)
rule \Cite{Wallach48}, and the lightness of the darker
region will be determined by the luminance ratio with the lighter
region.  Lightness will be constant until the relative size of the
dark region grows bigger than the relative size of the lighter
region: the \emph{area rule} applies and perceptual changes are produced.
Once the darker region is bigger, it appears lighter and lighter,
until, according to the highest luminance rule, it is anchored
at white (as it approaches $100\%$ size).  However, what happens
with the lighter region? At first, as the dark region is perceived
lighter, it remains at white.  Thus, a compression of lightness occurs,
despite of the luminance ratio being held constant.  Gradually,
however, the white region gets ``whiter than white'' (or super
white, or fluorent).  Finally, as the dark region approaches
$100\%$ and thus white, the white region is ``forced to relinquish
its white appearance and take on the appearance of self-luminosity''
(see \cite{GilchristEtAl99}, p.803).  However, as admitted by
Gilchrist and colleagues, their findings apply only to simple
Ganzfeld displays, and yet needs to be studied with more complex
displays (p. 802).\\
Because anchoring is related to lightness and thus to surface
representations, anchoring is not considered by the gradient
system in its present version.   Consequently, no area rule
applies to the gradient system.  The present results
suggest that gradient representation in the absence of concomitant
surface representations accounts for the perception of luminosity.
Note that a luminosity threshold is revealed as a function
of the luminance of the central square (\fig[CentralSquareVaries]).
The location of the threshold does not depend on the spatial frequency
of the carrier (that is, on the size of the central square).
However, the luminosity threshold of the gradient system does
not depend on the luminance of the other squares in the display,
and therefore is different from the luminosity threshold (i.e.,
the factor $1.7$) measured by \cite{BonatoGilchrist1994}.\\
Furthermore, it is not clear how the area rule applies to the
open glow display of \fig[GlareParade]: the apparent glow area
is increased with respect to the glow display, but this does 
not seem to compromise the perception of self-luminosity.\\
The observation that the lighter region appears
self-luminous if it is sufficiently small with respect to
the dark region could in principle be explained with the
formation of luminance gradients at the retina (e.g.,
``halos'', \cite{ZavagnoCaputo2005})
Such gradients may be produced at small and bright stimuli
embedded in a darker background due to increased pupil size
and the major part of the retinal array being adapted to
the darker background \Cite{Simpson1953,BettelheimPaunovic1979}.\\
The computational model of \cite{GrossHong03,GrossHong04,GrossHong06}
treats the generation of surface representations in the context of the
anchoring theory of lightness perception \Cite{GilchristEtAl99}.
Surface representations are generated by filling-in mechanisms.
Anchoring of perceived reflectance follows a modification
of the HLAW rule \Cite{Wallach48},
which is the  ``\emph{Blurred-Luminance-As-White-Rule}'' (BHLAW).
The modification overcomes problems like that ``a point-like
small bright patch on the visual field will be dealt with
the same as a large whiteboard occupying most of the visual
field''.  Thus, instead of looking simply for the highest
luminance value in an image and anchoring it at white, the
BHLAW rule suggests to anchor the highest value in a low-pass
filtered version of the image at white (where with ``image''
a filled-in surface representations is meant).  The
perception of luminosity occurs when an image region
which has the highest filled-in activity is smaller
than the size of the blurring kernel.  This mechanism
for producing luminosity effects is therefore different
from the gradient system, because it again measures
luminosity on the lightness scale.  Luminosity 
effects were demonstrated by the BHLAW-model with the
``Double Brilliant Illusion'' \Cite{BressanMingollaSpillmannWatanabe1997},
which creates a sensation of glow by using luminance
gradients analogously to the glare effect display.\\
The BHLAW model's overall behavior follows the anchoring theory
and area rule as described above.  An important difference between
the BHLAW model  and the gradient system is that the
former uses various resolution levels or scales for the
filling process (although not for the blurring kernel
for implementing the anchoring process).  
%
\begin{figure}[h!]
	\begin{center}
		\scalebox{0.30}{\includegraphics{\pics/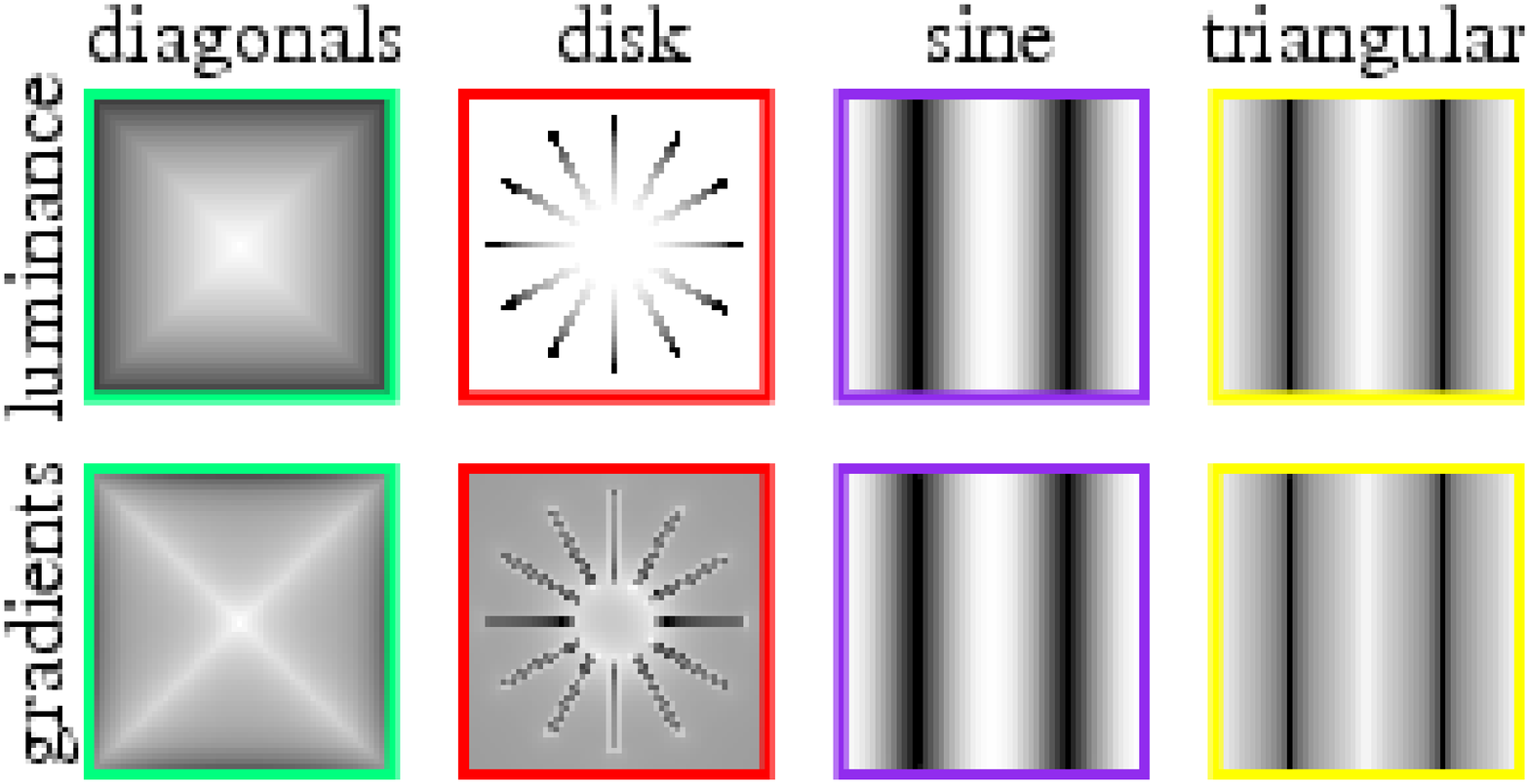}}
	\end{center}
	\Caption[MoreExamples][More examples for luminance gradients][{%
	The second row shows the gradient representations obtained with the images
	from the first row.  From left to right: the glowing diagonals of a luminance
	pyramid are predicted by the gradient system.  The brightness enhancement
	of an Ehrenstein Disk with an overlaid luminance gradient is predicted.
	A sine wave grating as an example of a nonlinear luminance gradient
	(doesn't the white stripe in the middle appear to glow?).  A triangular-shaped
	luminance profile reveals bright and dark Mach bands effects (horizontal
	stripes), which are also predicted by the gradient system.}]
\end{figure}
%
\subsection{Conclusions}
Recent psychophysical data concerning the luminosity effect suggest that luminance gradients are
a perceptual feature just like, for example, orientation, contrast or color \Cite{CorreaniSamuelLeonards06}.
Accordingly, in the present paper, a recently proposed theory about the processing of luminance gradients
has been evaluated in the context of luminosity perception.  The gradient system suggests how
luminance gradients are processed by the visual system at an early level, and how they can give
rise to perception of luminosity.  As gradient representations are thought to be complementary
to surface representations (i.e., lightness computations), possible interactions between both representations were discussed.
Although the gradient system is already successful at explaining several brightness illusions
in terms of luminance gradients (see also \fig[MoreExamples]), mechanisms which address the
interactions with surface representations and texture information need to be incorporated.
However, the precise nature of such mechanisms have yet to be established by corresponding
studies in fields like neurophysiology or psychophysics.\\
So, why are items displayed on a (light-emitting) computer screen are not perceived
as self-luminous? The answer is that there are no luminance gradients created by the
(light-emitting) pixels which could trigger gradient representations.  Only surface
representations are produced, and thus the displayed items are perceived
on the lightness scale.
%
%
\section*{Acknowledgments}
This work was supported by the \emph{Juan de la Cierva} program of the Spanish government
(BKC-IYK-6707), and the the MCyT grant SEJ 2006-15095.  The author likes to thank Fred Kingdom
and two reviewers from \emph{Vision Research} for their help in improving the first drafts.
The present preprint is an extended version of the ``official'' \emph{Vision Research}
manuscript.

\end{document}